\documentclass[11pt,a4paper]{article}
\pdfoutput=1
\usepackage{jheppub}
\usepackage{amsmath,amscd}
\usepackage{amsfonts}
\usepackage{amssymb}
\usepackage{graphicx}
\usepackage{color}
\usepackage[normalem]{ulem}
\usepackage{hyperref}
\usepackage{enumerate}
\usepackage{mathtools}
\usepackage{slashed}
\usepackage[symbol]{footmisc}

{

\newcommand{\ZZ}{\mathbb{Z}} 
\newcommand{\NN}{\mathbb{N}} 

\def\tr         {{\rm  tr}}

\def\calb         {{\cal B}}

\def\calf         {{\cal F}}

\def\calh         {{\cal H}}

\def\caln         {{\cal N}}
\def\calo         {{\cal O}}

\def\calq         {{\cal Q}}

\def\be{\begin{equation}}
\def\ee{\end{equation}}
\def\bea{\begin{eqnarray}}
\def\eea{\end{eqnarray}}

\def\a{\alpha}
\def\b{\beta}

\def\g{\gamma}

\def\d{\delta}
\def\e{\epsilon}

\def\l{\lambda}
\def\L{\Lambda}
\def\k{\kappa}
\def\f{\phi}

\def\m{\mu}
\def\x{\xi}
\def\n{\nu}
\def\o{\omega}
\def\O{\Omega}
\def\p{\pi}

\def\s{\sigma}

\def\t{\tau}

\def\sF{{{ F}\!\!\!\!\hskip.8pt\hbox{\raise1pt\hbox{/}}\,}}
\def\som{{{ \omega}\!\!\!\!\hskip.8pt\hbox{\raise1pt\hbox{/}}\,}}
\def\sJ{{{\rm J}\!\!\!\!\hskip.8pt\hbox{\raise1pt\hbox{/}}\,}}

\def\F{\Phi}
\def\pa{\partial}

\def\to{\rightarrow}
\def\nonu{\nonumber \\{}}
\def\half{{1 \over 2}}

\title{On conformal symmetry in large-$N$ quiver mechanics}

\author[a]{Andrea De Marco,}
\author[b]{Mustafa Mullahasanoglu,}
\author[a]{Joris Raeymaekers,}
\author[a]{Paolo Rossi}
\author[b]{and Gizem Seng\"or}

\affiliation[a]{CEICO, Institute of Physics of the Czech Academy of Sciences,\\  Na Slovance 2, 182 21 Prague 8, Czech Republic.}
\affiliation[b]{Physics Department, Boğaziçi University\\
		34342 Bebek / Istanbul, T\"urkiye.}
\emailAdd{marco@fzu.cz,mustafa.mullahasanoglu@std.bogazici.edu.tr,joris@fzu.cz,\\rossip@fzu.cz,gizem.sengor@bogazici.edu.tr}

\abstract{The microscopic description of extremal supersymmetric black holes in \linebreak  AdS$_2$/CFT$_1$ holography
has remained elusive despite recent  progress in the statistical   description of near-extremal black hole physics.	
In this work we revisit Denef's quiver mechanics description of D-brane bound states in the Coulomb branch, which displays an emergent conformal symmetry  in the  AdS$_2$ scaling limit. This conformal symmetry is however broken by superpotential corrections  
	near the locus where the Coulomb and Higgs branches meet, and its significance has so far remained unclear. In order to to clarify this issue, we derive and interpret  a fixed-point formula for the superconformal quiver index using localization techniques.
	Focusing on cyclic abelian quivers, we 
	  show that, in a certain large-$N$ limit (with  the rank $N$ of the quiver gauge group), the fixed points  are located  in the  regime where the conformal description is reliable. In this limit, our expression for the superconformal index  precisely captures a contribution to the microscopic scaling BPS index derived by Beaujard, Mondal and Pioline, which was hitherto not visible on the Coulomb branch. Our results are hoped to provide  a step towards a stringy realization of AdS$_2$/CFT$_1$ duality. }
\arxivnumber{}
\begin{document}
 \maketitle

\section{Introduction}
String theory has achieved  impressive success in  microstate counting for supersymmetric black holes \cite{Strominger:1996sh, Maldacena:1997de}. Since these computations typically rely on supersymmetry protection to perform the counting at weak gravitational coupling, they don't provide an explicit description of the microstates in the regime where gravity is turned on. Tracking the microstates to the regime where gravitational backreaction is significant has proven to be a hard problem, with various proposed approaches such   as the fuzzball programme (see \cite{Mathur:2005zp} for a review) or, more recently, the criterion of fortuity \cite{Chang:2024zqi}.

On general grounds, one would expect to be able to identify BPS black hole microstates in a strongly coupled  quantum mechanical theory under AdS$_2$/CFT$_1$ duality. The status of (pure) AdS$_2$/CFT$_1$ duality is however less clear than   its higher dimensional cousins. Since AdS$_2$ does not support any finite-energy excitations \cite{Maldacena:1998uz} and BPS black holes are expected to have gap separating  the ground states from the excited ones, one arrives at the picture, most notably advocated by Sen  \cite{Sen:2008vm},  that the AdS$_2$ decoupling limit should be dual to a collection of ground states, whose detailed  description 
has however remained elusive. 
We should remark that, while recent developments have shown that there is no  macroscopic  ground state degeneracy in   nonsupersymmetric  extremal black holes \cite{Iliesiu:2020qvm}, the picture of a large number of degenerate ground states  remains  plausible for  BPS black holes \cite{Heydeman:2020hhw}.

While recent years have witnessed great progress in understanding the physics of near-extremal black holes, specifically through JT gravity (and its random matrix completion) and SYK-type models (see \cite{Mertens:2022irh} resp.  \cite{Chowdhury:2021qpy} for reviews), these have neither
 confirmed nor rejected Sen's proposal for the microscopics of supersymmetric pure AdS$_2$. In the case of JT gravity, the AdS$_2$ entropy is an input parameter, i.e. the constant dilaton VEV around which one expands, and is not explained microscopically by the model.  SYK models and their supersymmetric cousins on the other hand do predict a macroscopic entropy if the large-$N$ limit is taken before the zero-temperature limit. The statistical nature of these models however indicates that they should be viewed as  effective coarse-grained descriptions which remain agnostic about the precise microscopic Hilbert space. It remains an important open question whether   an embedding in string theory can provide a more fundamental understanding of the AdS$_2$ ground states.

In this work we revisit a tried-and-tested microscopic description of D-brane dynamics in $\caln=2$ string compactifications 
to 3+1 dimensions in the form of Denef's 
quiver quantum mechanics \cite{Denef:2002ru}.
This description captures the  stringy low-energy interactions between D-brane centers
and  has proven to correctly capture the    BPS spectrum  including subtle wall-crossing effects \cite{Denef:2007vg}. 
In particular, the Coulomb branch description of the theory, which we will focus on in this work,  is typically valid  close to the supergravity regime and  `knows' about some of the supergravity  BPS  equations and quantum numbers. Furthermore, for D-brane systems allowing for an AdS$_2$ scaling limit, the quiver Coulomb description formally displays an emergent superconformal symmetry \cite{Bena:2012hf,Anninos:2013nra,Mirfendereski:2020rrk}, making it tempting to propose it as (a sector of\footnote{One should keep in mind that a black hole of a given total charge can be `deconstructed' into systems of individual D-branes in a large number of ways.  Each such quiver realization  therefore describes only a sector of the tentative full CFT$_1$.}) the tentative CFT$_1$ description. However, there are two closely related caveats to such an interpretation. 

Firstly, one might object that the Coulomb branch approximation is likely to miss the degrees of freedom relevant for single-center black holes and AdS$_2$ physics. Indeed, it is well-known that the Coulomb branch approximation is blind to  a class of BPS states which arise as middle-cohomology classes in the Higgs branch \cite{Bena:2012hf}. These  `pure-Higgs' states are angular momentum singlets and are highly  numerous in a certain large-charge limit; therefore they are widely  believed to represent single-center black hole microstates. At the level of the refined Witten index of the theory  
Manschot, Pioline and Sen \cite{Manschot:2010qz,Manschot:2011xc, Manschot:2012rx} have shown that the naive Coulomb branch result  needs additional corrections  to account for scaling configurations, which include the missing pure-Higgs states. More recently, Beaujard, Mondal and Pioline \cite{Beaujard:2021fsk} have found a suggestive formula, based on the Jeffery-Kirwan residue formula due to Hori and Yi \cite{Hori:2014tda}, for these scaling contributions which we will compare to in our work. 
The reason why the analysis in the Coulomb branch approximation misses states is presumably a breakdown of its validity due to the integrated chiral fields becoming massless,   which in particular results in the  decompactification of  Coulomb branch moduli space  in the AdS$_2$ scaling regime.  
 This is in fact closely related to the appearance of conformal symmetry, which implies that the Hamiltonian $H$ 
in the AdS$_2$ regime  has a continuous spectrum extending down to zero, making the Witten index a highly subtle  object. However, the presence of conformal symmetry also allows us in this case to define a different, superconformal, index which roughly speaking instead employs  $L_0$ as the `Hamiltonian'. Since $L_0$ has discrete spectrum in these models, we expect the superconformal index to be  better defined object. Our main technical result will be to derive a fixed-point formula for the quiver superconformal index\footnote{Inspiring works by Dorey and collaborators have computed superconformal indices in models with more supersymmetry,  formulated in terms of different multiplets \cite{Barns-Graham:2018xdd,Dorey:2018klg,Dorey:2019kaf}.}  using localization techniques. Like the Manschot-Pioline-Sen fixed-point formula  for the Witten index \cite{Manschot:2010qz,Manschot:2011xc, Manschot:2012rx}, it receives
 contributions from collinear D-brane configurations, where the centers obey equations which are a deformation of  
 Denef's scaling supergravity equations which has the effect of  regularizing the solution space.

A further caveat in the conformal Coulomb branch approach is that the conformal limit of the    theory was obtained from the full quiver theory  upon integrating out the chiral multiplets at the 1-loop level. This approximation 
neglects corrections at two loops coming from superpotential terms \cite{Anninos:2013nra}, 
which provide  `relevant'  corrections  which break conformal symmetry near the locus where the Coulomb and Higgs branches meet. However, focusing on abelian cyclic quivers with rank $N$ gauge group, we will argue that in a certain large $N$
limit  these conformal-breaking corrections can be neglected\footnote{A different approach was taken in \cite{Anninos:2016szt} (see also \cite{Biggs:2023mfn}), where the theory was considered in the Higgs branch upon averaging over the superpotential coefficients. This was shown to also  lead to an emergent conformal symmetry, albeit of a different kind than in our context (i.e. with different scaling dimensions of the fields), and in a different large-$N$ limit.}.  In this limit, our expression for the superconformal index  precisely captures a contribution to the aforementioned  microscopic scaling BPS index formula of by Beaujard, Mondal and Pioline \cite{Beaujard:2021fsk}. This contribution, which was hitherto not visible in the Coulomb branch approximation,   
suggests that  conformal symmetry has a role to play  in large-$N$ quiver mechanics and,  it might be hoped, provides   a step towards a stringy realization of AdS$_2$/CFT$_1$ holography.

This paper is structured as follows. In Section \ref{Secquiver} we review the basics of Coulomb branch quiver mechanics. In Section \ref{Secconf} we discuss the AdS$_2$ deep scaling limit which leads to an emergent $D(2,1;0)$ superconformal symmetry in the quiver theory, as well as some expected corrections to this picture. Section \ref{Secreps} is devoted to basic representation theory of the $D(2,1;0)$ algebra as well as the definition of a superconformal index which contains information on the short multiplet spectrum. Section \ref{Secloc} contains our main technical result in the form of a fixed-point formula for the quiver superconformal index derived using localization methods. We also rederive  the known localization formula for the standard Witten index as a check of our approach. In Section \ref{Seceval} we discuss the evaluation of our fixed-point formula for abelian cyclic quivers and identify a certain large-$N$ limit where corrections to the conformal approximation are negligible. In this limit we can identify our  superconformal index formula with a contribution to the microscopic scaling BPS index of \cite{Beaujard:2021fsk}.
We end in Section \ref{Secdisc} with some open questions raised by our work and avenues for future research.

\section{Quiver quantum mechanics in the Coulomb branch} \label{Secquiver}
Quiver quantum  mechanics \cite{Denef:2002ru} is an $\caln =4$ supersymmetric theory describing 
the low-energy dynamics of  D-branes wrapped on Calabi-Yau cycles in type II string theories. 
  It is obtained by reducing the quiver gauge theory on the brane worldvolumes to the 0+1 dimensional worldlines the branes trace out in the noncompact dimensions. The Lagrangian contains interacting vector multiplets
and chiral multiplets 
and is determined by the quiver data. We will focus on the Coulomb branch approximation, which arises from integrating out the chiral multiplets at one loop, which is justified only if they are sufficiently massive, i.e. the VEVs of the vector multiplet scalars is sufficiently large and the branes are far apart. We will return to the issue of the validity of the Coulomb branch approximation  in Section \ref{Seccorrs}. 

\subsection{Quiver Lagrangian in the Coulomb branch}
We will restrict our attention to abelian quivers where each node or D-brane center carries a $U(1)$ gauge group, which amounts  to restricting the D-brane centers to carry primitive charges. For   each D-brane center   the theory contains an $\caln=4$ vector multiplet whose bosonic content\footnote{In addition, the vector multiplet contains a worldline gauge potential $C_t$ which however decouples for the abelian quivers under consideration.}  consists of three spatial coordinates ${x_i},i = 1, 2 ,3$, representing the spatial position of the brane and an auxiliary field $D$.
 The fermionic superpartners are labelled as\footnote{Our fermions are related to those in \cite{Mirfendereski:2020rrk} as
 $
 	 \l_\a^{\  +} = - i \l_\a, \  \l^{\a -} = - i \bar \l^\a .
 	 $}
 $\l^{\a \tilde \b},\  \a, \tilde \b  \in \{+, -\}$. The index structure reflects  that they transform as doublets under two separate $su(2)$ $R$-symmetries as we shall see in detail below.  The fermions satisfy the
 reality conditions
 \be 
 (\l^{\a \tilde \b})^* = \a \tilde \b \l^{ -\a, -\tilde \b}
 \ee
Our conventions for spinor index contractions are somewhat different from those  in \cite{Denef:2002ru, Mirfendereski:2020rrk} and are spelled out in Appendix \ref{Appferm}.

Let us now discuss the Coulomb branch Lagrangian in more detail. In what follows we will work in string units where
\be 
l_s :=  \sqrt{2 \p \a '} \equiv 1.
\ee
We label the D-brane centers and vector multiplets by an index $a = 1 , \ldots N$.
The center of mass dynamics is universal and decouples from the
dynamics of the relative coordinates and their superpartners. The relative Lagrangian depends only on the differences $ \F_{ab} := \F_a - \F_b$ for $\F \in (x, \l, \bar \l, D)$ and  reads
\bea
L &=&L^{(0)} + 
L^{(1)} + L^{(2)}\label{Lag1}\\
L^{(0)} &=&  - f_a  D^a \nonu
L^{(1)} &=&  - U_a (x)  D^a+  A_{ia} (x) \dot x^{ia} -{1 \over 2} \pa_{ia} U_b  (x) \l^a \s_i \l^b\nonu
L^{(2)} &=& \half G_{ab} (x)  \left(\dot x^{ia} \dot x^{ib} + D^a D^b - i \l^a \dot \l^b 
\right) + {1 \over 4}  \pa_{ic}  G_{ab} (x)  \left(   \l^a\s_i \l^b   D^c+ \e_{ijk}  \l^a \s_j \l^b \dot x^{kc} \right)\nonu 
&& + {1 \over 32}  \pa_{jc}\pa_{jd}G_{ab}  (x) (\l^a \tilde \s_i \l^b  ) (\l^c \tilde \s_i \l^d).
\label{Lagcomp}
\eea
Though not manifest in the way we have written it, the coefficients in this Lagrangian are such that it actually only depends on the relative fields $\F_{ab}$.
This is due to properties such as
\be 
f_a v^a= U_a v^a = A_{ia} v^a= G_{ab} v^a =0,
\ee 
where $v^a$ is the vector $v^a = (1,1,\ldots , 1)$. One way to restrict to a manifestly non-redundant description is to set $x_{iN} = \l_{N a \tilde \b} = D^N =0 $ and restrict the indices $a,b, \ldots $ to the range $1, \ldots , N-1$.
 
Each of the terms $L^{(0)},L^{(1)},L^{(2)}$  is $\caln=4$ supersymmetric by itself. The term $L^{(0)}$ of zeroth order in  time derivatives of the bosonic fields contains the Fayet-Iliopoulos constants $f_a$ which depend on the asymptotic moduli of the Calabi-Yau compactification. The  first order part  $L^{(1)}$  captures   electromagnetic-type interactions between the centers: the $U_a$ are electrostatic Coulomb potentials and the $A_{ia}$ Dirac monopole vector potentials:
\bea  
A (x) &=& - \sum_{a<b} \k_{ab} A_i^D (x_{ab} )dx^i_{ab}, \qquad 
A^D (x) = {x^1 dx^2 - x^2 dx^1\over 2 r (z \pm r)}\\
U_a (x)& =& \sum_{b, b\neq a} \frac{\kappa_{ab}}{2r_{ab}}.\label{Ux}
\eea
(the  Dirac potential with positive/negative sign is in a gauge which is regular on the positive/negative $z$-axis). The integer coefficient $\k_{ab}$ is the Dirac-Schwinger-Zwanziger (DSZ) pairing between the   charges of brane $a$ and brane $b$. When $\k_{ab}$ is nonzero the $a$ and $b$ branes carry mutually nonlocal charges and can be thought of as electron-monopole pair with respect to some $U(1)$  gauge field.                                                                                          
The electromagnetic fields satisfy
\be
\pa_{ia} U_{b} = \pa_{ib} U_{a}  =\e_{ijk} \pa_{jb} A_{ka}\label{Nis4UA}
\ee
which is the condition for $L^{(1)} $ to be invariant under $\caln=4$ supersymmetry.

The term $L^{(2)}$ containing two time derivatives of the bosons describes a supersymmetric nonlinear  sigma model with target space metric $G_{ab} (x)$.
The condition for $\caln=4$ invariance of  $L^{(2)} $ imposes that $G_{ab}$ is the Hessian of a function $\calh$ (the `Hesse potential'), i.e.
\be
G_{ab} = \half\pa_{ia}\pa_{ib}  \calh, \qquad 
 \pa_{ia}\pa_{jb} \calh = \pa_{ib}\pa_{ja} \calh ,
 \ee
 For  quiver mechanics the Hesse potential and metric are given by\footnote{We note that $\calh$ is not unique and can be redefined by adding a function with vanishing   Hessian matrix.}
\bea 
\calh &=&- \sum_{a, b , a\neq b}
\left({\m_{ab} r_{ab}^2\over 6}  +\frac{|\kappa_{ab}|}{4r_{ab}}\log r_{ab}\right)\nonu
G_{ab} &=& \m_{ab}-\frac{|\kappa_{ab}|}{4r_{ab}^3}+ \delta_{ab}\sum_{c,c\neq {a}}\left( \frac{|\kappa_{ac}|}{4r_{ac}^3}- \m_{ac}\right) .\label{HGnonscaling}
 \eea
 where $\m_{ab} =  \m_{ba}$ are the reduced masses of the brane pairs.

In addition to experiencing  electromagnetic fields and a curved geometry, the D-brane centers also move in a potential, which can be obtained from integrating out the auxiliary fields $D^a$ which leads to  
\be
V = \frac{1}{2} G^{ab}(f_a+U_a)(f_b+U_b).
\ee

\subsection{Symmetries and Noether charges}
The Lagrangian (\ref{Lagcomp}) preserves $\caln =4$  supersymmetry as well as an  
R-symmetry algebra $ su(2)_J \oplus \widetilde{su(2)}$ under which the supercharges transform in the $(j,\tilde j) =( 1/2,  1/2)$ representation. The $su(2)_J$ algebra corresponds to the  angular momentum of the D-brane system under spatial rotations, while  $\widetilde{su(2)}$ acts only on the fermions and descends from the R-symmetry of the centrally extended $\caln=2$ superalgebra in $3+1$ dimensions. The  symmetry generators\footnote{The supercharges in the present $(1/2, 1/2)$ notation are related to the charges $Q_\a, \bar Q^\a$ in \cite{Mirfendereski:2020rrk} as $Q^{\a+} = -i \bar Q^\a/\sqrt{2}, Q^{\a -} =  -i  Q^\a/\sqrt{2} $.}  are explicitly given by
\bea
J_i &=& \e_{ijk}x^a_j \left( p_{ka}- A_{ka}  \right)+ U_a x^a_i + s_i, \qquad s_i := -{1 \over 4} \l^a \s_i \l_a\\ 
\tilde J_i 
&=& -{1 \over 4} \l^a \tilde  \s_i \l_a \\
 Q^{\a\tilde \b} &=&- {1 \over \sqrt{2}} \left((f_a + U_a) \l^{a\a\tilde \b} + i\left( p_{ia} - A_{ia} \right) (\l^a \s_i)^{\a \tilde \b}  \right)\label{Qs}\\
 H &=&\half G^{ab} (P_{ia} P_{ib} - D_a D_b) + \half \pa_{ia}U_b \l^a \s_i \l^b \\
&+& (f_a + U_a - {1 \over 4} \pa_{ia} G_{bc} \l^b \s_i \l^c) D^a-  {1 \over 32}  \pa_{jc}\pa_{jd}G_{ab}  (x) (\l^a \tilde \s_i \l^b  ) (\l^c \tilde \s_i \l^d)\\
P_{ia}&:=& G_{ab} \dot x_i^b = p_{ia} - A_{ia} -{1 \over 4}\e_{ijk} \pa_{ja} G_{bc} \l^b \s_k \l^c.
\eea 
 The $su(2)_J$ angular momentum generators $J_i$ contain a spin part $s_i$ and an orbital part;
as usual these are not separately conserved due to the  magnetic field term in $L^{(1)}$ and the spin-orbit coupling terms in $L^{(2)}$. 
The supercharges obey the reality conditions
\be 
(Q^{\a \tilde \b})^* = \a \tilde \b Q^{ -\a, -\tilde \b}.
\ee 

To find the canonical brackets between the fields, one observes that the fermions are subject to second class constraints and can be dealt with using the Dirac bracket. A standard analysis, spelled out in Appendix \ref{AppDB}, leads to  the following canonical brackets:
 \be 
 \{ x^a_i, p_{bj} \}_{\rm PB} = \d^a_b \d_{ij}, \qquad  \{\l_a^{\a \tilde \a} , \l^{b \b \tilde \b} \}_{\rm PB} = - i \d^b_a \e^{\a\b}\tilde \e^{\tilde \a \tilde \b} \qquad  \{ p_{ia}, \l_b^{ \a \tilde \a} \}_{\rm PB} = \half\pa_{ia} G_{bc} \l^{c \a \tilde \a} \label{DB}
 \ee
 Under these, the charges obey the following Poisson bracket algebra
\begin{align} 
\{J_i, J_j \}_{\rm PB}&= \e_{ijk} J_k, & \{\tilde J_i, \tilde J_j \}_{\rm PB}&= \e_{ijk} \tilde J_k \label{Nis41}\\
i \{J_i, Q^{\a \tilde \a}\}_{\rm PB} &= \half ( \s_i Q) ^{\a \tilde \a} 
,& i \{ \tilde J_i, Q^{\a \tilde \a}\}_{\rm PB} &= \half  (\tilde \s_i Q)^{\a \tilde \a} \\
i \{ Q^{\a \tilde \a}, Q^{\b \tilde \b}\}_{\rm PB} &=   \e^{\a \b} \e^{\tilde \a \tilde \b} H .&&\label{Nis43}
\end{align}

The supersymmetry   transformations of the fields $\d_\x \F :=\{ \x_{\a \tilde \b}  Q^{\a \tilde \b}, \F \}_{\rm PB}$ read\footnote{The transformation parameters $\x_{\a \tilde \b}$ are related to those in \cite{Mirfendereski:2020rrk} as $\x_\a = -{i \over \sqrt{2}} \x_{\a -}, \bar \x^\a =   -{i \over \sqrt{2}} \x_+^{\a}$. }
	\be 
			\d_\x x_i^a =- {i \over \sqrt{2}} \x \s_i \l^a ,\qquad
			\d_\x D^a =   {1 \over \sqrt{2}} \x \dot \l^a, \qquad
		\d_\x \l^a = {1 \over \sqrt{2}}  ( -\dot x_i^a \s_i \x + i D^a \x ).
		\ee
		The classical moduli space  of supersymmetric solutions consists of constant bosonic configurations,
$\dot x_{i}^a=\l^a  =0$, 
satisfying the D-term equations
\be 
D_a = f_a +  \sum_{b, b\neq a} \frac{\kappa_{ab}}{2r_{ab}}=0.\label{Denefeqs}
\ee
These equations, which constrain the positions of the D-brane centers,  also appear as integrability conditions for the BPS equations describing the backreacted  multi-centered  supergravity solutions \cite{Denef:2000nb,Bates:2003vx}, and are referred to as the Denef equations. 
Furthermore, on classical ground states the 
angular momentum  Noether charges reduce to 
		\be 
		\left.J_i\right|_{\rm vac} =  U_a x^a_i =\half \sum_{a,b, a<b} {\k_{ab} \over r_{ab}} x^{ab}_i,
		\ee
		which is precisely  the expression for the ADM  angular momentum of the corresponding  supergravity solution \cite{Denef:2000nb}.

\subsection{Underlying wHKT geometry}
We end this section with a comment on a more geometric sigma model reformulation of these models which goes under the name of automorphic duality, see \cite{Delduc:2006yp, Mirfendereski:2020rrk} for more details. For a general review of supersymmetric sigma models, see \cite{Smilga:2020nte}.  By introducing additional fields $x_0^a$ and making the substitution
\be 
D^a \to D^a - \dot x_0^a
\ee 
we obtain a classically equivalent model with gauge symmetries allowing to set $x_0^a$ to be constant and return to the original model.  In the new formulation, the bosonic fields $x^a_\m = (x_0^a, x_i^a)$ are on the same footing\footnote{The variables $(x_\m, \l_\m)$ are said to form a $(4,4,0)$ multiplet, whereas the original variables $(x_i, \l^{\a \tilde \b}, D)$ comprise a $(3,4,1)$ multiplet.}   as their superpartners $\l_\m^a$, while $D^a$ play the role of worldline gauge fields which gauging a translation symmetry of the target space. 
The resulting $4(N-1)$ dimensional target  manifold has special geometric properties: the metric defined by the sigma model is weakly  hyper-K\"ahler with torsion (wHKT), while the  
field strengths
 \be 
 \calf_{ab}:= \left(\begin{array}{cc} \pa_{ia }  U_b    &  F_{ia jb}   \\0 & - \pa_{ia }  U_b    \end{array} \right)
 \ee
 are selfdual in 4D for all $a,b$.  The generalization of the current work to more general gauged sigma models with wHKT  target spaces  
 will be discussed in a separate publication \cite{nextpaper}.

\section{Deep scaling limit and superconformal invariance}\label{Secconf}
In this Section we discuss the emergence of $D(2,1;0)$ superconformal symmetry for scaling quiver systems,  corresponding to the formation of an AdS$_2$ throat region in the corresponding supergravity solutions \cite{Anninos:2013nra, Mirfendereski:2020rrk}. We also review how this symmetry is broken by superpotential effects near the locus where the Coulomb and Higgs branches meet. The study of conformal symmetry in quantum mechanics goes back to \cite{deAlfaro:1976vlx}, while for discussions of superconformal mechanics we refer to \cite{Michelson:1999zf, Fedoruk:2011aa} and references therein. 
\subsection{Scaling charges and deep scaling limit}
For generic D-brane charges and DSZ products $\k_{ab}$, the Coulomb branch moduli space carved out by the D-term constraints is compact and quantization of the system poses no significant obstructions. 
As we will discuss in more detail in the next Section, a  supersymmetric Witten-type index can then  be defined and computed reliably using localization. 

This picture however changes drastically when the DSZ products are such that the Denef equations allow 
for scaling solutions. Let us illustrate this for a 3-center abelian quiver. The scaling regime occurs when the intersection products satisfy triangle inequalities:
\be 
0<\k_{12}< \k_{23} + \k_{31}, \qquad 0<\k_{23}< \k_{31} + \k_{12}, \qquad 0<\k_{31}< \k_{12} + \k_{23}.
\ee
In this regime, the Denef equations (\ref{Denefeqs}) allow for a perturbative solution in a small parameter $\e$ where the coordinate distances between the centers become very small:
\be 
r_{ab} = \e |\k_{ab} | + \calo (\e^2),\label{scalingsols}
\ee
where the subleading parts are adjusted to satisfy the  equations (\ref{Denefeqs}). 
For more than 3-centers, similar scaling solutions appear for example in cyclic abelian quivers (see Figure \ref{Figcyclicquiver}) obeying the  inequalities
\be 
0<\k_{12}< \k_{23} + \ldots + \k_{N1} \qquad {\rm and \ cyclic }.\label{Nineqs}
\ee
In the corresponding multi-center supergravity solution, the D-brane centers can be seen to disappear down a deep AdS$_2$ throat in the limit $\e \to 0$ \cite{Denef:2007vg}. From the point of view of  the Coulomb branch mechanics,  a noncompact runaway direction in moduli space  opens up  in the scaling regime where  $r_{ab} \to 0$, as can be seen from the moduli space metric (\ref{HGnonscaling}).
This noncompactness complicates the computation of the Witten index \cite{Manschot:2010qz,Manschot:2011xc,Kim:2011sc}. 
As we shall presently see, these subtleties in the scaling regime   are closely  related to the emergence of conformal symmetry. 

Let us consider a `deep scaling limit' of the theory which singles out the $\e \to 0$ regime of (\ref{scalingsols}). In particular, we scale the time and field variables as follows:
\begin{equation}
t\rightarrow \e^{-1} t\,,\quad x^{ia}\rightarrow \e x^{ia}\,,\quad \lambda_a^{\alpha\tilde \b} \rightarrow \e^{3/2} \lambda_a^{\alpha\tilde \b}\,,\quad D^a\rightarrow \e^2 D^a\label{confscal}
\end{equation}
and take $\e \to 0$.
The Denef equations reduce in this limit to
\be 
\sum_{b, b\neq a} \frac{\kappa_{ab}}{2r_{ab}}=0
\ee
and become insensitive to the FI parameters $f_a$ and therefore to 
 the Calabi-Yau moduli. One also shows that the   ground states classically have vanishing angular momentum, $\left.J_i\right|_{\rm vac} = 0$.
 The moduli-independence and vanishing angular momentum are  properties which the deep scaling solutions share with  single-center BPS black holes, suggesting they might play the role of  semiclassical microstates. At the quantum level, we expect  the scaling limit (\ref{confscal}) to accurately
 describe states in which
 \be 
 \langle r_{ab} \rangle \ll f_a^{-1}, \qquad \forall a,b,\  a \neq b.
 \ee

 At the level of the Coulomb branch Lagrangian (\ref{Lagcomp}), under the scalings  (\ref{confscal})  the couplings $f_a$ and $\m_{ab}$ are `irrelevant'   in the sense that 
 \be 
 f_a \to \e f_a, \qquad \m_{ab} \to \e^3 \m_{ab} ,
 \ee
 and therefore these couplings disappear from the Lagrangian. The remaining moduli space metric in the $\e \to 0$ limit takes the form\footnote{ We should note that, despite first appearances, the metric is not diagonal as it is expressed in terms of redundant variables ${\bf x}_{ab}$. 
 	When expressed in terms of  $3(N-1)$ non-redundant coordinates, such as ${\bf x}_{a,a+1}^i, a = 1, \ldots, N-1$ it is no longer diagonal.}
 \be 
ds^2 = \sum_{a,b,a<b} {|\k_{ab}| \over 4 r_{ab}^3} d {\bf x}_{ab} \cdot  d {\bf x}_{ab}\label{confmetr}
 \ee
It can be rewritten as a cone over a compact base manifold $\calb$ as
\be 
ds^2 = dr^2 + r^2 ds^2_\calb,
\ee
where the radial coordinate is given by
\be 
r^2 =  \sum_{a,b,a<b} {|\k_{ab} | \over r_{ab}} 
\ee
The UV-like limit  of small $r_{ab}$ where all the centers approach each other  is, from the point of view of the Coulomb branch mechanics, an IR limit  
$r \to \infty$  which is at infinite distance and 
where the proper  volume diverges. 

\subsection{Emergent superconformal symmetry}
 One can show that in the deep scaling limit (\ref{confscal}) the Lagrangian  becomes conformally invariant \cite{Anninos:2013nra,Mirfendereski:2020rrk}, thanks to the properties
\bea
x^{ic}\partial_{ic}G_{ab}&=&-3G_{ab}, \qquad G_{ab} x^{ib} = - \pa_{ia} \left( G_{ab}x^{ja}x^{jb} \right),\label{hompropG} \\
x^{ib}\partial_{ib}U_a&=&-U_a, \qquad x^{jb}\partial_{jb}A_{ia}=x^{jb}\partial_{ia}A_{jb}.
\eea
One finds for the  dilatation and special conformal  charges
\bea 
D &=&  x^a_i p_{ai} - {3 i \over 4} \l^a   \l_a \\
K&=&  
  =  \sum_{a,b,a<b} {|\k_{ab} | \over 2 r_{ab}} \left( = { r^2 \over 2}\right)
\eea
These (and other charges) below are given at an initial time slice $t=0$: since they don't Poisson-commute with the Hamiltonian, additional  $t$-dependent terms are generated at different times.

The conformal generators $H,D,K$ should fit together with the $\caln = 4$ supercharges and $su(2) \oplus \widetilde{su(2)}$ R-symmetry into a superconformal algebra. According to the classification of superconformal algebras \cite{Nahm:1977tg, Claus:1998us}  this algebra should belong to the 1-parameter family of algebras $D(2,1; \a)$. It was established in \cite{Mirfendereski:2020rrk} that the relevant algebra  is in fact
$D(2,1;0)$, which is a semidirect sum of the superconformal algebra $psu(1,1|2)$ and the $\widetilde{su(2)}$ part of the R-symmetry. This algebra contains four conformal supercharges $S^{\a\tilde \b}$ obtained as $ \{ Q^{\a \tilde \b}, K \}_{\rm PB} = S^{\a \tilde \b}  $:
\be \label{Ss}
S^{\a  \tilde \b} ={i \over \sqrt{2}}\pa_{ia }K (\l^a \s_i)^{\a \tilde \b}.
\ee
It will be convenient to work in the super-Virasoro-like basis 
\be
 L_0 = \half ( H+K) , \qquad L_{\pm 1} =  \half ( H-K) \pm i D, \qquad 
 G_{\pm \half}^{\a \tilde \b} =    Q^{\a \tilde \b} \pm i  S^{\a \tilde \b} \label{Gs}
\ee
in which the conformal  supercharges obey the reality conditions
\be 
\left(G^{\a \tilde \b}_{\pm \half}\right)^* = \a \tilde \b G^{ -\a, -\tilde \b}_{\mp \half}.
\ee 
and obey the basic Poisson anticommutators
\bea 
i \{G^{\alpha \tilde \alpha}_{\pm \half},G^{\beta \tilde \beta}_{\pm \half}\}_{\rm PB} &=&  2 \epsilon^{\alpha \beta} \epsilon^{ \tilde \alpha  \tilde \beta}L_{\pm 1}    \\
i \{G^{\alpha  \tilde \alpha}_{\half},G^{\beta  \tilde \beta}_{-\half}\}_{\rm PB} &=& 2 \epsilon^{\alpha \beta} \epsilon^{ \tilde \alpha  \tilde \beta} L_0 +  2 \epsilon^{ \tilde \alpha  \tilde \beta} J_i \s_i^{\a \b}  
\eea
 The superconformal transformations of the fields $ \d_{\g^\pm} \F :=  \{\g^\pm_{\a \tilde \b} G^{\a \tilde \b}_{\pm \half}, \F \}_{\rm PB} $ are 
  \begin{align} 
 \d_{\g^\pm} x_i^a &= -{i \over \sqrt{2}} \g^\pm \s_i \l^a, &  \d_{\g^\pm} D^a &=   {1 \over \sqrt{2}} \g^\pm (\dot \l^a \pm 3 i \l^a)\\
 \d_{\g^\pm} \l^a & = {1 \over \sqrt{2}}  \left(- ( \dot x_i^a\pm 2 i x^a_i)  \s_i \g^\pm  + i D^a \g^\pm  \right). && \label{deltaG}
 \end{align}

 \subsection{Superpotential corrections}\label{Seccorrs}
 In the above we saw that the quiver Coulomb Lagrangian (\ref{Lagcomp}) has a smooth  limit under the scaling (\ref{confscal}),  where the limiting theory becomes superconformally invariant. We should however keep in mind that 
  (\ref{Lagcomp}) was obtained by integrating out the chiral multiplets in the one-loop approximation  and keeping the terms up to second order in time derivatives.
  However, when the quiver contains closed loops, as is always the case for scaling quivers,  it is possible to add a gauge-invariant superpotential for the chiral multiplets which leads to   corrections to the Coulomb branch Lagrangian.  
 Since the superpotential terms in the full  quiver Lagrangian (see \cite{Denef:2002ru}) do not directly couple vector- and chiral multiplets, they start correcting the Coulomb branch Lagrangian at two loops. 
 The general form of the corrections is discussed in \cite{Anninos:2013nra} to which we refer for more details.
 
  While the precise superpotential coefficients are  hard to compute directly from string theory in the current setting of type II  Calabi-Yau compactifications, it was argued in \cite{Denef:2007vg} that a superpotential is present and can  be assumed to be cubic.  We write it in the schematic form
  \be 
  W  = w \F^3,
  \ee
  where the $\F$ are chiral multiplets and $w$ denote coupling constants which we assume to be of order one in our string units. 
   While the first-order Lagrangian $L^{(1)}$ can be argued to be one-loop exact, the second order Lagrangian $L^{(2)}$ will receive corrections. Dimensional analysis shows that a correction term to the sigma model  metric of the form
  \be 
  \d_W L^{(2)} = w^2 {\dot {\bf x}^2 \over r^6}.\label{Wcorr}
  \ee
  can (and does \cite{Anninos:2013nra}) arise. From this we see that $w$ is a relevant coupling  under the scaling (\ref{confscal}),
  \be 
  w \to \e^{-{3\over 2}} w.
  	\ee
 The correction term (\ref{Wcorr}) breaks superconformal invariance\footnote{It's interesting to note that it would be invariant if we assign  a different scaling dimension to the fields, namely ${\bf x}_{ab} \to \e^{1/4} {\bf x}_{ab}$. This scaling would be consistent with $D(2,1;3)$ superconformal invariance \cite{Mirfendereski:2022omg} rather than $D(2,1;0)$. The meaning of this  and the possible relation to the results of \cite{Anninos:2016szt} remains  to be explored.}  	and  dominates over the $D(2,1;0)$ invariant terms (\ref{confmetr}) for sufficiently small  $\langle r_{ab} \rangle$. The conformally invariant description should be reliable as long as, in string units,  
 \be 
  \langle r_{ab} \rangle \gg w^{2\over 3}\sim \calo (1).
  \ee
  In Section \ref{SeclargeNsols} below we will see how to realize this condition in a certain large-$N$ limit. 
 
 \section{(Super-)conformal representation theory and indices}\label{Secreps}
 In this Section we collect some results on representation theory and generalized  Witten indices. We will discuss both the standard $\caln =4$ supersymmetry applicable to quiver models away form the conformal regime, and the $D(2,1;0)$ superconformal symmetry relevant for conformal quivers. This will set the stage for our localization computation of both the standard Witten index (which we can compare with known results) and the $D(2,1;0)$ superconformal index in quiver theories, and presents an opportunity to review and compare the different indices considered in the literature. 
 \subsection{$\caln=4$ supersymmetry and index}
 Let us start with theories possessing $\caln =4$ supersymmetry and R-symmetry $su(2)_J \oplus \widetilde{su(2)}$. The symmetry algebra is the quantum version\footnote{Our notation does not distinguish between quantum operators and classical Noether charges; we hope that it is clear from the context which is meant.}  of (\ref{Nis41}-\ref{Nis43}):
 \bea 
\,  [J_i, J_j] &=& i \e_{ijk} J_k,\qquad  [\tilde J_i,\tilde J_j] = i \e_{ijk} \tilde J_k\\
\,  [J_i, Q^{\a \tilde \a}] &=& \half (\s_i Q )^{ \a \tilde \a },\qquad  [\tilde J_i, Q^{\a \tilde \a}] = \half (\tilde \s_i  Q )^{\a \tilde \a}  \\
\,  \{ Q^{ \a \tilde \a} ,  Q^{ \b \tilde \b} \} &=& \e^{\a\b} \e^{\tilde \a \tilde \b} H
  \eea
Long representations are built on a $(j,\tilde j )$ multiplet of states with energy $E \neq 0$ and contain $16 (2j+1)(2 \tilde j+1)$ states which carry the following R-symmetry representations
  \bea 
& 4  (j, \tilde j)_E \oplus  2(j + \half, \tilde j+1/2)_E  \oplus  2 (j + \half, \tilde j-1/2)_E \oplus  2 (j - \half, \tilde j+1/2)_E \oplus  2 (j- \half, \tilde j-1/2)_E&\nonu
& \oplus  (j + 1, \tilde j )_E   \oplus  (j , \tilde j + 1)_E  \oplus  (j -1 , \tilde j )_E \oplus  (j , \tilde j-1 )_E &
  \eea
  A short multiplet consists of a  $(j,\tilde j )_0$ multiplet of states with energy $E = 0$ and contains $1/16$  of the  states of a analogous long multiplet.
 
 Information about the short multiplet content is contained in a refined Witten index which is simply the standard Witten index with respect to a $\caln=2$ subalgebra. To construct it, we pick  a complex supercharge, say $Q^{-+}$, and its Hermitean conjugate $(Q^{-+})^\dagger = - Q^{+-}$. We observe that these charges commute with the combination $J_3 + \tilde J_3$, which allows us to write a refined Witten index  
  \be
   I[q] := \tr (-1)^{2 \tilde J_3} e^{- \b H} q^{- 2( J_3 + \tilde J_3)}.\label{WittenI}
   \ee
 It receives contributions only from short multiplets, with  a multiplet $(j, \tilde j)_0$ contributing
   $$
   (-1)^{2 \tilde j} \chi_j (q) \chi_{\tilde j} (q).
$$
Therefore we can write
\be 
 I[q] = \sum_{j, \tilde j} (-1)^{2 \tilde j} N_0 (j, \tilde j) \chi_j (q) \chi_{\tilde j} (q).
 \ee
 where $ N_0 (j, \tilde j)$ is the number of short multiplets in the $(j, \tilde j)$ representation. 
   
   Let us comment on the relation of the Witten index (\ref{WittenI}) to other indices and some concrete expectations for quiver models 
   \begin{itemize}
   	\item The index (\ref{WittenI}) is defined with respect to the parity operator $(-1)^{2 \tilde J_3}$, which in quiver models plays the role of worldline fermion parity. Often one considers instead an  index $\O[q]$ where the parity is replaced by $(-1)^{2 J_3}$ which in quiver models is interpreted as the spacetime fermion parity:
   \be 
   \O[q] := \tr (-1)^{2 J_3} e^{- \b H} q^{ 2( J_3 + \tilde J_3)} 
   \ee
   (in keeping with literature conventions  we have also replaced $q \to q^{-1}$).
    This alternative index is referred to as the `protected spin character', see \cite{Gaiotto:2010be,Lee:2012naa} for details.  The indices $I[q]$ and $\O [ q]$ are simply related as  
   	\be 
   	\O[q] :=  I[-q^{-1}]
   	\ee
   	\item For quiver mechanics,  extensive study of both Higgs and Coulomb branches \cite{deBoer:2008zn,deBoer:2009un,Bena:2012hf,Manschot:2010qz,Manschot:2011xc,Lee:2012sc,Manschot:2012rx,Lee:2012naa,Hori:2014tda} has shown  that supersymmetric ground states  occur in only two types of R-symmetry representations: either  singlets under $\widetilde{su(2)}$ or singlets under $su(2)$. These two types have different physical interpretations (see Table \ref{Tablemults}). 
   	\begin{table}	\begin{center}
   		\begin{tabular}{|c|c|c|c|c| c|} \hline
   			type & {$ su(2)_J \times \widetilde{su(2)}$} & interpretation &$   I[q]$& $  \tilde I [q]$\\ \hline  \hline 
   			Coulomb &  {$ (j,0)$} & 	\begin{tabular}{ll}multi-center \\ bound states  \\ \ \end{tabular} & $\chi_j (q) $  
   			&  $q^{2  j}$ \\ \hline 
   			pure-Higgs &   {$ (0, \tilde j)$}  & 	\begin{tabular}{ll}single-center BH \\microstates\\ \   \end{tabular} &$(-1)^{2 \tilde j} \chi_{\tilde j}(q) $  & $(-1)^{2 \tilde j} \chi_{\tilde j} (q) $ \\
   			\hline
   			\end{tabular}\end{center}\caption{Supersymmetric ground states with various $R$-charges occurring in quiver quantum mechanics and their contribution to the Witten index $I[q]$. For comparison, we also list the contributions of similar R-charged  short multiplets  to  the superconformal index $\tilde I[q]$.}\label{Tablemults}
   		\end{table}
   	The multiplets of the form $(j,0)_0$  are interpreted as bound states of multicenter D-brane systems with mutually nonlocal charges, as these 
   	always carry angular momentum stored in crossed electric and magnetic fields. These short multiplets are visible both in the Higgs and Coulomb branch, where the appropriate description depends on the string coupling \cite{Denef:2002ru}. On the other hand,  scaling quivers can also have ground states which are angular momentum singlets of the type $(0, \tilde j)_0$. These are commonly interpreted as 	single-center black hole microstates, firstly because MSW-type \cite{Maldacena:1997de}  single-center black holes do not carry angular momentum and secondly because detailed counting studies have shown them to be highly  numerous in a large-charge limit. These states were found only in the Higgs branch \cite{Bena:2012hf} and are therefore often called `pure-Higgs' states. 
   	\item Manschot, Pioline and Sen \cite{Manschot:2010qz,Manschot:2011xc,Manschot:2012rx} have also considered a different quantity, namely the ground state character
   		\be 
   	\O_{\rm MPS}[q] := \lim_{\b \to \infty} \tr(-1)^{2 J_3} e^{- \b H} q^{ 2 J_3} \label{OmMPS}
   	\ee
   	Since none of the supercharges commute with $J_3$, this object is generically not an index and the projection on ground states has to be imposed by by hand by taking the zero-temperature limit. However for quiver systems $\O_{\rm MPS}[q]$ was argued \cite{Lee:2012sc,Lee:2012naa,Manschot:2012rx} to be  an index as well: the contributions of Coulomb-type states $(j,0)_0$ to $\O_{\rm MPS}[q]$  and to  the protected spin character $\O[q]$ are identical, while pure-Higgs contribute differently to $\O[q]$ and $\O_{\rm MPS}[q]$   but have been argued to be robust under variations of the moduli and string coupling due to their interpretation as single-center black hole microstates.
  \end{itemize}
 \subsection{$D(2,1;0)$ superconformal symmetry and index}\label{Secind}
 We now turn to systems with $D(2,1;0)$ superconformal invariance.  The algebra is
 \bea
 \, [L_m,L_n] &=& (m-n)L_{m+n}\\
 \, [L_m, G_{A}^{\a \tilde \a}] &=& \half (m - 2  A)  G_{m+ A}^{\a \tilde \a}\label{L0G}\\
 \, [J_i, G_{A}^{\a \tilde \a}] &=&  \half ( \s_i G_{A}) ^{\a \tilde \a},\qquad  [\tilde J_i, G_{A}^{\a \tilde \a}] = \half ( \tilde \s_i G_{A})^{\a \tilde \a} \label{JG}\\
 \{G^{\alpha \tilde \alpha}_{\pm 1/2},G^{\beta \tilde \beta}_{\pm 1/2}\} &=& 2\epsilon^{\alpha \beta} \epsilon^{ \tilde \alpha  \tilde \beta}L_{\pm 1}   \label{d21aalg1} \\
 \{G^{\alpha  \tilde \alpha}_{1/2},G^{\beta  \tilde \beta}_{-1/2}\} &=&2 \epsilon^{\alpha \beta} \epsilon^{ \tilde \alpha  \tilde \beta} L_0 +  2\epsilon^{ \tilde \alpha  \tilde \beta}  \s_i^{\, \a \b} J_i  \label{d21aalg4} 
 \eea
 We will assume an inner product under which the  generators obey  the Hermiticity properties
 \bea
 L_m^\dagger &=& L_{-m}, \qquad J_i ^\dagger = J_i, \qquad \tilde J_i ^\dagger =\tilde J_i\\
 \left( G_{\pm \half}^{\a\tilde \a} \right)^\dagger &=& \a \tilde \a  G_{\mp \half}^{-\a\, - \tilde \a}.\label{HermGs}
 \eea

 A generic long multiplet $(j,\tilde j)_{h_0}$ is built on a $(j,\tilde j)$ R-symmetry  multiplet of primary states with $L_0$-weight $h_0$, which are annihilated by all $L_0$-lowering operators ($L_1$ and  all four $G_{1/2}^{\a \tilde \b}, \a, \tilde \b = \pm$). 
 Acting with the $L_0$-raising operators $G_{-1/2}^{\a \tilde \b}, \a, \tilde \b = \pm$ and R-symmetries generically yields states up to $L_0$-level 2 which decompose into 16 $su(2)_J \oplus \widetilde{su(2)}$ irreps as follows:
 \be
 \begin{array}{|c|l|c|}\hline
{\rm level}& 	L_0 & {\rm R-symmetry\  irreps}\\ \hline 
0& 	h_0 & (j , \tilde j)\\ 
 \half&	h_0 + \half & (j+ \half , \tilde j + \half)\oplus(j+ \half , \tilde j - \half)\oplus(j- \half , \tilde j + \half)\oplus(j- \half , \tilde j - \half)\\
1& 	h_0 + 1 & (j+ 1 , \tilde j  )\oplus(j  , \tilde j+ 1)\oplus2 (j  , \tilde j )\oplus(j-1 , \tilde j )\oplus(j , \tilde j-1 )\\
 {3 \over 2}&	h_0 + {3 \over 2} & (j+ \half , \tilde j + \half)\oplus(j+ \half , \tilde j - \half)\oplus(j- \half , \tilde j + \half)\oplus(j- \half , \tilde j - \half)\\
2& 	h_0+ 2 & (j , \tilde j)\\ \hline
 \end{array}\label{longmult}
 \ee
 In the above table,  the highest weight states of the R-symmetry multiplets at higher levels are typically nontrivial linear combinations carrying the appropriate weights. This will be discussed in more detail in \cite{nextpaper}.
  The full (infinite-dimensional) representation is obtained by acting in addition with the raising operator $L_{-1}$ on these states. 

 Unitarity restricts the range  of allowed $h_0$ for given $j,\tilde j$. 
 Consider the anticommutators
 \be 
 \left\{ G_{-\half}^{\a \tilde \b} , \left(G_{-\half}^{\a \tilde \b}\right)^\dagger \right\} = 2L_0 - 2\a J_3 \label{anticommunit}
 \ee
 for $\a, \tilde \b \in \{,+,-\}$.
 Unitarity requires the right-hand side to be positive for all states in the multiplet. This leads to the unitarity bound
 \be 
 h_0 \geq j.
 \ee 
 When this bound is saturated, the multiplet shortens  compared to the generic situation indicated in  (\ref{longmult}) and contains `chiral primary' states which are annihilated by additional fermionic generators. It is straightforward to see that, at level 0, the states of lowest $su(2)$ weight $j_3 = -j$ must be annihilated by the $L_0$-raising operators $G_{-1/2}^{- \tilde \b}$ (in addition to the lowering operators $G_{1/2}^{\a \tilde \b}$). These states form a spin$-\tilde j$ multiplet under $\widetilde{su(2)}$. Similarly, the states with highest  $su(2)$ weight $j_3 = j$ are annihilated by the $L_0$-raising operators $G_{-1/2}^{+ \tilde \b}$. The structure of the short multiplet $(j, \tilde j)_j$ is as follows:
 \be
\begin{array}{|c|l|c|}\hline
	{\rm level}& 	L_0 & {\rm R-symmetry\  irreps}\\ \hline 
0& 	j & (j , \tilde j)\\ 
\half& 	j + \half & (j- \half , \tilde j - \half)\oplus(j- \half , \tilde j + \half)\\
 1&	j + 1 & (j -1 , \tilde j )\\
 \hline
 \end{array}\label{shortg0}
 \ee
 When we let the weight $h_0$ of a long multiplet approach $j$, it decomposes into four short multiplets:
 \be
   ( j,\tilde j)_{h_0} \to  ( j, \tilde j)_j \oplus \left(j+ 1/2, \tilde j- 1/2\right)_{ j+ 1/2}\oplus \left(  j- 1/2, \tilde j- 1/2\right)_{j- 1/2}  \oplus \left( j , \tilde j+ 1\right)_j. 
 	\ee

To define a superconformal index receiving contributions only from chiral primaries, we proceed in a similar manner as above   and pick a subalgebra generated by a complex superconformal charge and its conjugate and write a Witten-like index as in $\caln =2$ theories. In what follows we will pick $ G_{-1/2}^{-+}$
and $ (G_{-1/2}^{-+})^\dagger= - G_{1/2}^{+-}$ and  observe that they commute with $J_3 + \tilde J_3$  so that we can define a refined superconformal index $\tilde I [q]$ as
\be 
\tilde I [q]= \tr (-1)^{2 \tilde J_3}e^{ -\b \tilde H } q^{ - 2( J_3 + \tilde J_3)}.\label{Itildedef} 
\ee
Here, the role of the `Hamiltonian' is played by
\be 
  \tilde H :=   \left\{ G_{-\half}^{-+}, ( G_{-\half}^{-+})^\dagger\right\}
   = H + K + 2 J_3 
  \ee
The index  $\tilde I [q]$ receives contributions from chiral primary states satisfying 
\be 
G_{-1/2}^{-\tilde \a}|CP \rangle = (G_{-1/2}^{-\tilde \a})^\dagger |CP \rangle=0.
\ee    	
 	Similarly one could introduce an index receiving contributions from `anti-chiral  primaries' annihilated by $G_{-1/2}^{+\tilde \a}$  and $(G_{-1/2}^{+\tilde \a})^\dagger$ by flipping the sign in front  of $J_3$ in $\tilde H$; this object  captures the same representation content, as we have seen that each short multiplet contains both `chiral' and `anti-chiral' states related by a Weyl reflection on the R-symmetry weights.
 	
 From the above considerations we find that 	the contribution of a short multiplet $( j , \tilde j)_j$ to the index is
 	\be
	\tilde I_{( j , \tilde j)_j}[q] =	(-1)^{2 \tilde j} q^{2 j} \chi_{\tilde j} (q).\label{multcontrsc}
 	\ee
 Letting $N_S(j, \tilde)$ denote the number of short multiplets in a given theory, the index evaluates to: 
 	\be  
 			\tilde I[q] =  \sum_{j, \tilde j} (-1)^{2 \tilde j} N_{S} (j,\tilde j)   q^{ 2 j} \chi_{\tilde j} (q) 
 	\label{inddegs}
 	\ee
 	The coefficients in a Laurent expansion
 	\be
 	\tilde I [q] = \sum_{n \in \ZZ} \tilde a_n q^n
 	\ee
 	contain the following information on the short multiplet degeneracies
 	\bea 
 	\tilde a_n 
 	= \sum_{2 j \in \NN} \sum_{2 \tilde j \geq |n-2j|} (-1)^{2\tilde j} N_S (j ,\tilde j).
 	\label{Laurentdegs}
 	\eea
 	Let us now comment on some properties of  the  superconformal index and its relation to indices defined  in the literature.
 	\begin{itemize}
 		\item Unlike the Witten-type indices discussed above, the superconformal index is generally not symmetric\footnote{ Also, as we will see in the examples, it is typically not  a Laurent polynomial.} under sending $q \leftrightarrow q^{-1}$. Indeed,  
 		we observe from (\ref{multcontrsc})  that  poles  and constant terms in the Laurent expansion of   $\tilde I[q]$ around $q=0$ can only arise from multiplets with $\tilde j \geq j$; in particular pure-Higgs-like multiplets 
 		 $(0,\tilde j)_0$  give rise to poles or constant terms, while Coulomb-like multiplets $(j,0)_j$  contribute positive powers of $q$.
 			In Table \ref{Tablemults} we have listed for comparison the contribution of Coulomb-like   and pure-Higgs-like  multiplets to $\tilde I[q]$. 
 		\item As for the Witten index, we can define an alternative superconformal index $ \tilde \O [q]$ weighted by the spacetime fermion parity  $ (-1)^{2 J_3}$ (an sending $q \to q^{-1}$ for convenience). The two  choices are related as \be \tilde \O [q] := \tr (-1)^{2 J_3}e^{ -\b \tilde H } q^{  2( J_3 + \tilde J_3)} = \tilde I[-q^{-1}]\label{Omtdef}
 		\ee
 		\item The index $\tilde \O[q]$ agrees with the superconformal index discussed in \cite{Gaiotto:2004pc}. Indeed, 
 		defining
 		\be 
 		y = e^{-2 \b (1-\g)} q^{2}, \qquad z = e^{-2 \b \g} q^{-2},
 		\ee
 		we can rewrite (\ref{Omtdef}) as
 		\be
 		\tilde \O[q] =\tr (-1)^{2 J_3} y^{L_0 - \a J_3} z^{L_0 - \tilde \b \tilde J_3}\label{Istrom}
 		\ee
 		which  agrees with (A.3) in \cite{Gaiotto:2004pc}. 
 		\end{itemize}
     
 \section{Fixed-point index  formulas from localization}\label{Secloc}
 In this Section we derive the fixed-point formula for the superconformal quiver index which forms the main technical result of this work. We will discuss a formal relation between the superconformal and Witten indices which allows us to compute both quantities simultaneously. 
 Since the latter was already computed in the literature using other methods this will provide a consistency  check of our approach.
 In an upcoming publication \cite{nextpaper}  we will generalize these localization formulas to systems with more general $D(2,1;\a )$ symmetry and provide explicit checks in simple systems where the short multiplet spectrum can be obtained explicitly.
Fixed-point formulas for models with $\caln =2$ superconformal symmetry were derived and  checked in concrete examples in \cite{Raeymaekers:2024usy,Raeymaekers:2024ics}. Superconformal indices were also studied  in models with more supersymmetry,  formulated in terms of different multiplets, in \cite{Barns-Graham:2018xdd,Dorey:2018klg,Dorey:2019kaf}.  
 \subsection{Preliminary observations}
 Before starting the localization computation, we make two useful observations. 
 Our goal is to compute the superconformal index, which can be written as
 \be 	\tilde I [q ] = \tr (-1)^{2 \tilde J_3}e^{ -\b \tilde H} q^{- 2(J_3 + \tilde J_3)},
 \ee
 where $\tilde H$ is the `Hamiltonian'
\be \tilde  H = H + \o^2 K +2 \o J_3 .\label{tildeHom}\ee
Here we have introduced a positive parameter $\o$ which arises from an inner automorphism of the algebra by conjugating all generators with $ e^{ \ln \o D} $ and an additional rescaling of the inverse temperature $\b$. Clearly,  $\o$ is a spurious parameter in the sense that any strictly positive value is equivalent. However, in the limit $\o \to 0$ we obtain a different object which is the Witten index,
\be 
``\lim_{\o \to 0} \tilde I [q ]  = I [q ]".
\ee 
This equation should be interpreted as taking $\o \to 0$ in the integrand of the path-integral before integrating.
 The effect of turning on the $\o$-dependent terms in (\ref{tildeHom})  can in fact be described in a  simple manner. We first observe that the $\caln=4$ supercharges and the superconformal charges are related by similarity transformations involving the special conformal generator $K$ in the following way:
 \be 
 G_{\pm \half}^{\a \tilde \a} =\sqrt{ \o  } e^{ \mp  \o K}  Q^{\a\tilde \a}  e^{ \pm \o K}.
 \ee
 In particular, to define the superconformal index we picked  the superconformal charge $ G_{- \half}^{-+}$ and to define the Witten index we picked  the supercharge $  Q^{-+}$. Writing out the explicit expressions (\ref{Qs}) and (\ref{Ss}, \ref{Gs}) we find\footnote{In deriving this, it is important that the shifts of  $U_a$ and $A_{ia}$ are real and the same for $G_{- \half}^{-+}$ and $(G_{- \half}^{-+})^\dagger$, so that the Hermiticity relation is preserved.} that the above similarity transformation amounts to a shift of the electrostatic potentials and the magnetic vector potentials as follows:
 \be 
 G_{- \half}^{-+}  = \left.   Q^{-+} \right|_{ U \to \tilde U, A \to \tilde A}  , \qquad   (G_{- \half}^{-+})^\dagger = \left.   (Q^{-+})^\dagger \right|_{ U \to \tilde U, A \to \tilde A},
 \ee
 where
 \be 
\tilde U_a = U_a- \o \pa_{3a}K, \qquad  \tilde A_{ia} = A_{ia} -\o  \e_{ij3} \pa_{ja} K . \label{UtAt}
\ee
Similarly, one checks that the `Hamiltonian' $\tilde H $ in (\ref{tildeHom})  is related to the original one $H$ by the same shifts, and in addition a shift of the auxiliary field: 
\be 
  \tilde D_a = D_ a - \o \pa_{3a} K .\label{Dt}
\ee

The main ingredient in  localization computations is the identification of a supersymmetry-exact term that can be added  to the action  with a large coefficient without consequence,  preferably chosen such that the path integral localizes  to a sum over isolated fixed points. To describe our choice, let us first consider the Lagrangian $\tilde L$ corresponding to the `Hamiltonian' $\tilde H$ in (\ref{tildeHom}). 
It will be convenient to work with new fermionic variables  	which transform irreducibly under the diagonal 
$su(2)_{\rm diag}$ symmetry generated by
\be  J_i^{\rm diag} = J_i + \tilde J_i.
\ee
The fermions $\l^{\a \tilde \b}$, which form a bispinor under $su(2) \oplus \widetilde{su(2)}$,  decompose into a triplet $\l^i$  and a singlet $\l^0$ under $su(2)_{\rm diag}$. We introduce the projections on these representations as follows
\be \l^i = d^i_{\a \tilde \b} \l^{\a \tilde \b} \qquad  \l^0 = d^0_{\a \tilde \b} \l^{\a \tilde \b} , \ee
where for definiteness we take
\begin{align}
d^i_{\a \tilde \b} &
 = {1 \over \sqrt{2}} \left( \left(
\begin{array}{cc}
-1 & 0 \\
0 & 1 \\
\end{array}
\right), \left(
\begin{array}{cc}
i & 0 \\
0 & i \\
\end{array}
\right),\left(
\begin{array}{cc}
0 & 1 \\
1 & 0 \\
\end{array}
\right)\right), & d^i_{\a \tilde \b} &
=  {1 \over \sqrt{2}} \left(
\begin{array}{cc}
0 & -1 \\
1 & 0 \\
\end{array}
\right).
\end{align}
 In terms of the new variables the first-order Lagrangian $L^{(1)}$ takes the form\footnote{One needs the identity
	\be - i \l_i^{(a} \l_j^{b)} +  \e_{ijk} \l_k^{(a} \l_0^{b)}= {1 \over 2} \e_{ijk} \l^a \s_k \l^b.\ee}
\be 
L^{(1)} = - U_a D^a + A_{ia} \dot x^a_i - \pa_{ia} U_b \l_i^a \l_0^b + {i\over 2}  F_{ia\ jb} \l^a_i  \l_j^b.\label{L1diag}
\ee
From the  discussion around (\ref{UtAt} ,\ref{Dt}) we know that  $\tilde L$ is obtained from the original Lagrangian (\ref{Lagcomp}) by the same shifts of $U_a$ and $A_{ia}$ and  the auxiliary fields $D^a$, i.e.
\bea
\tilde L &=& \tilde L ^{(1)} + L^{(2)}\\
\tilde L ^{(1)}   &=& - \tilde U_a \tilde D^a + \tilde A_{ia} \dot x^a_i - \pa_{ia} \tilde U_b \l_i^a \chi^b + {i\over 2} \tilde F_{ia\ jb} \l^a_i \l_j^b.\label{L1t}
\eea 
We should note that $\tilde U_a$ and $\tilde A_{ia}$ do {\em not} satisfy the condition (\ref{Nis4UA}) required for $\caln=4$ invariance: the new Lagrangian breaks the symmetry down to an $\caln=2$ superconformal subalgebra  generated by $ G^{-+}_{-1/2}$ and 
$ G^{+-}_{1/2}$. Let us consider  the real superconformal charge
\be 
\tilde \calq := G^{-+}_{-1/2} -  G^{+-}_{1/2}.
\ee 
One finds\footnote{We should note that these  transformations are not simply given by (\ref{deltaG}) which were appropriate for the original Lagrangian  $L^{(1)}$. While the phase space supercharges  are still given by (\ref{Qs},\ref{Ss},\ref{Gs}), the relation between $p_{ia}$ and $\dot x^{ia}$ is different for $\tilde L^{(1)}$ than it was for $ L^{(1)})$, leading to  This leads to an effective shift of $\dot x^{ia}$ in the RHS of the transformation law (\ref{deltaG}),   leading to (\ref{tildeQtransfor}).}  that it generates the transformations
\begin{align}
\d_{\tilde \calq} x_i^a &= {i }  \l_i^a, & \d_{\tilde \calq} \tilde D^a &=   \dot \l_0^a \\
\d_{\tilde \calq} \l_i^a &=   \dot x_i^a &
\d_{\tilde \calq} \l_0^a &= {i }\tilde D^a  
 \label{tildeQtransfor} 
\end{align}
Note that the ${\tilde \calq}$-symmetry acts on $\tilde L$ essentially in the same way the $\calq = Q^{-+} - Q^{+-} $-symmetry acts on the original Lagrangian $L$.
Now it is easy to verify that $\tilde L^{(1)}$ is ${\tilde \calq}$-exact:
\be 
\tilde L^{(1)} = \d_{\tilde \calq}\left(    i \tilde U_a \l_0^a + \tilde A_{ia} \l_i^a\right).
\ee  
This will provide us with the susy-exact term in our localization computation which we now turn to. 

\subsection{Fixed-point formula for superconformal index}
Let us now derive a fixed-point formula for the superconformal index $\tilde I [q]$ from a  localization argument in the path integral. By taking the parameter $\o \to 0$ (or, equivalently, replacing $\tilde U_a \to U_a, \tilde A_{ia} \to  A_{ia}$) this will yield  an expression for the Witten index which can be compared with earlier derivations. Evaluating the index on a pure phase $q = e^{i \theta}$, we write it as
\be 
\tilde I[ e^{i \theta}] = \tr (-1)^{2 \tilde J_3} e^{ - \b \tilde H (\theta) }
\ee
where  $\tilde H (\theta)$ contains a perturbation by the  chemical potential $\theta$:
\bea 
\tilde H (\theta ) &=& H +\o^2 K + 2 \o J_3 +   {2 i \theta \over \b} ( J_3 + \tilde J_3) \label{tildeHtheta}\\
&=& \left( H + {2 i \theta \over \b} ( J_3 + \tilde J_3) \right)_{| U\to \tilde U, A \to \tilde A, D \to \tilde D} + {2 i \theta \over \b} K
\eea
The last term is due to the identity $(J_3)_{U,A,D} =  (J_3)_{| U\to \tilde U, A \to \tilde A, D \to \tilde D} + K$.  The corresponding Lagrangian has as its first-order part
\be 
\tilde L^{(1)}( \theta ) = \tilde L^{(1)} - { 2i \theta \over \b} ( \tilde U_a x^3_a + K  + i \e_{ij3} \l_i^a \l_{ja} )\label{tildeL1theta}
\ee
In this section we will  fix the translation invariances of the relative Lagrangian by  setting $x^i_N= \l_N^{\a \tilde \b}= D^N=0$ and letting $a$ run from 1 to $N-1$.
The action is invariant under a  supersymmetry  transformation  which is a deformation of (\ref{tildeQtransfor})
\begin{align}
\d_{\tilde \calq}  x_i^a &= i  \l_i^a & \d_{\tilde \calq}  \tilde D^a &= \dot \l_0^a\\
\d_{\tilde \calq}  \l_i^a &= \dot x_i + { 2i \theta \over \b} \e_{ij3} x_j^a &
\d_{\tilde \calq}  \l_0^a &= i \tilde  D^a\label{deltatildeQtheta}
\end{align}
The first-order Lagrangian  is almost susy-exact in the sense that
\be 
\tilde L^{(1)}( \theta ) = \d_{\tilde \calq} ( i \tilde U_a \chi^a+ \tilde  A_{ia} \l_i^a) -{2 i \theta \over \b}\left( U_a x_3^a - \e_{ij3} A_{ia} x_j^a + i \e_{ij3}\l_i^a \l_{ja} \right) 
\ee
We note that 
 the non-exact part in brackets is related to the Noether charge as $(J_3 + \tilde J_3)^{U,A}_{|p_{ia} \to 0} $. 

 The upshot is that we can write the  superconformal
 index as a  Euclidean path integral
 \bea
  \tilde I[e^{i \theta}] &=& \int [dx d\tilde D d\l_i d\l_0 ]_\b e^{ - \int_0^\b d\t \tilde L_E(\theta) }\\
\tilde  L_E(\theta) &=&   \left( \L L_E^{(1)}   + L_E^{(2)}\right)_{| \dot x_i^a \to  \dot x_i^a + {2 \theta \over \b } \e_{ij3}x_j^a}  + {2 i \theta \over \b} (U_a x_3^a - \e_{ij3} A_{ia}x_j^a + i \e_{ij3} \l_i^a \l_{ja})\nonu
 \tilde L_E^{(1)}  &=&  \tilde U_a \tilde D^a -i \tilde  A_{ia} \dot x^a_i + \pa_{ia} \tilde U_b \l_i^a \chi^b - {i\over 2} \tilde  F_{ia\ jb} \l^a_i \l_j^b.\label{indexPI}
 \eea
 Here, the path integrations run over field configurations which are periodic in Euclidean time with period $\b$. The parameter $\L$ multiplies a susy-exact part; we will localize the path integral by taking $\L\to \infty$.
 On general grounds, the path integral should localize on the supersymmetric locus of the transformations (\ref{deltatildeQtheta})\footnote{More general solutions to $\d \l_i^a =0$ are possible, but those would not be periodic in $\t$.}, namely
 \be 
 x_{1,2}^a (\t ) = 0, \qquad  x_{3}^a (\t)=  x_{3 0}^a, \qquad D^a (\t) = 0.
 \ee
In other words, the path integral localizes on collinear  configurations where all the D-branes are located on the $z$-axis.  We now expand the fields in the neighbourhood of the susy locus. It will be convenient to treat the constant modes of the fields $x_3^a, D^a, \l_3^a$ and $ \l_0^a$ on a separate footing as follows:
 \begin{align}
 	x_{1,2}^a (\t) &= 0 + {1 \over \sqrt{\L}} \d  x_{1,2}^a (\t)&  \l_{1,2}^a &= 0 + {1 \over \sqrt{\L}} \d  \l_{1,2}^a(\t) \\
 	x_{3}^a (\t)&= z^a + {1 \over \sqrt{\L}} \d \tilde  x_3^a (\t) & \l_{3}^a(\t)&= 0 + {1 \over \L}  (\d \l_3^a)_{0} + {1 \over \sqrt{\L}} \d \tilde  \l^a (\t)\\
 	D^a &= 0 + {1 \over \L}  \d D_{0}^a + {1 \over \sqrt{\L}} \d \tilde  D^a (\t) & 	\l_0^a &= 0  + (\d \l_0^a)_{0} + {1 \over \sqrt{\L}} \d \tilde  \chi^a (\t)
 \end{align}
 (here, a tilde denotes the non-constant fluctuation part). The first-order Lagrangian in (\ref{indexPI}) then reduces to
 \be 
 {\L L_E^{(1)}}_{| \dot x_i^a \to  \dot x_i^a + {2 \theta \over \b } \e_{ij3}x_j^a} =\tilde U_a (z) \d D_0^a + \pa_{3a} \tilde U_b (z)  (\d\l_3^a)_0 (\d \l_0^b)_{0}- {i \over 2} \d X  \cdot \tilde\calf (z) \cdot \calo  \cdot \d X -  {i \over 2} \d \Psi  \cdot \tilde\calf (z) \cdot  \d \Psi, 
 \ee
 where we introduced the short-hand notation
 \bea 
 \d X^a &=& ( \d x_1^a, \d x_2^a, \d \tilde x_3^a, \d \tilde D^a ), \qquad \d \Psi^a = ( \d \l_1^a, \d \l_2^a, \d \tilde \l_3^a, \d \tilde \l_0^a )\\
\tilde \calf_{\m a \n b} (z):&=& \left( \begin{array}{cc}\tilde F_{ia jb}(z) &  \pa_{ia }\tilde U_b (z) \\ - \pa_{ia }\tilde U_b  (z) &0 \end{array} \right), \qquad \m , \n = 1, \ldots , 4\\
 &=& \left( \begin{array}{cccc}0 &\tilde F_{1a2b} (z)&0\\ - \tilde F_{1a2b} (z)&0 &0&0\\0&0&0& \pa_{3a }\tilde U_b (z)\\0&0& - \pa_{3a } \tilde U_b  (z) &0 \end{array} \right)\\
 \calo_\theta & =&  \left( \begin{array}{cccc} \pa_\t &{2 \theta \over \b}&0&0\\- {2 \theta \over \b} &\pa_\t &0&0 \\
 0&0&0 &i\\
 0&0&-i&0
 \end{array} \right)
 \eea 
 We note that both $\calf (x_0 )$ and $\calo$ are antisymmetric operators which furthermore commute, so that the bosonic kinetic operator is symmetric. Performing the zero-mode and fluctuation integrations we find
 \be
 \tilde I[e^{i \theta}] = \int \prod_{a=1}^{N-1} dz^{a} (i)^{N-1} \d^{N-1} (\tilde  U_a (z) ) \det (\pa_{3b}\tilde  U_c (z) )
 e^{- 2 i \theta  J_3 (z) } \left( \det { \tilde \calf (z) \over \tilde \calf (z) \cdot \calo_\theta }\right)^{\half}.\label{localint}\\ 
 \ee
Here, the factor $ e^{- 2 i \theta  J_3 (z) }$ comes from the non-susy-exact part of the first-order action in (\ref{tildeL1theta}). We have also used that the second-order Lagrangian $ {L_E^{(2)}}_{| \dot x_i^a \to  \dot x_i^a + {2 \theta \over \b } \e_{ij3}x_j^a} $  vanishes on the susy locus. The delta-functions pick out the collinear configurations   which are solutions of $\tilde U_a (z_*)=0$; we assume (and will verify below) that these equations have only isolated solutions. 
We also see from (\ref{localint}) that our localization scheme is consistent only if the matrix $\tilde \calf $ is nonsingular at the fixed points, 
which reduces to the condition
\be \det \tilde F_{1a 2b}(z_* ) \neq 0 \qquad {\rm and } \qquad \det \pa_{3 a} \tilde U_b (z_* )\neq 0.\label{consistemcycond}\ee
 Working out the functional determinant in (\ref{localint}) gives
 \bea
 (\det \calo_\theta)^{- \half} &=& \left( \prod_{m \in \ZZ}{ (2 \p i n)^2 + (2 \theta )^2\over \b^2 } \prod_{n \in \ZZ \backslash \{0\}} (-1) \right)^{- {N-1 \over 2}}\\
 &=& \left( \left({2 \theta \over \b} \right)^2  \prod_{m>0}\left( { 2\p n \over \b}\right)^4
 \prod_{n>0}  \left(1 - \left( {\theta \over \p n}\right)^2 \right)^2\right)^{- {N-1 \over 2}}\\
 &=& (4 \sin^2 \theta )^{{1-N\over 2}}
 \eea 
  
 Putting everything together we find our fixed-point formula for the  superconformal index 
\be
\boxed{\tilde   I[q]  =  ( q - q^{-1})^{1-N} \sum_{z_*} {\rm sgn} \det   (\pa_{z^a}\pa_{z^b}\tilde  \F (z_*) )q^{ -2 J_3 (z_*)}, }\label{indfp}
\ee
 where
\be 
\boxed{J_3 (z_* ) = \half \sum_{a<b} \k_{ab} {\rm sgn} z_{ab}^* .}
\ee
The fixed points $z^*$ are solutions of
\be 
\boxed{    \sum_{b\neq a} \frac{\kappa_{ab}}{2|z_{ab}^*| } - \o \pa_{3a}K(z_*)= 0}\label{confdenefeqs}
 \ee
 We note that  the $\o$-dependent term yields a deformation (and, as we shall shortly see, a regularization)  of the collinear Denef equations in the deep scaling regime where $f_a \to 0$.
 These localization equations  can be derived from a  Coulomb-like potential $\tilde \F (z)$ as $\pa \tilde  \F / \pa z_{a}=0, a = 1, \ldots, N-1$, where
\be
\boxed{\tilde \F = \half \sum_{a<b} \k_{ab} {\rm sgn} z_{ab}  \ln  | z_{ab} | - \o K (z)}\label{tildeF}
\ee
We remark that the index (\ref{indfp}) depends only on the ordering of the centers on the $z$-axis, not on their precise positions. Therefore solving the fixed-point equations (\ref{confdenefeqs}) numerically  will suffice to get the exact index.

\subsection{ Witten index   formula}
Letting $\o \to 0$ in the above expressions, which amounts to replacing $\tilde U_a \to U_a$ and $\tilde A_{ia} \to A_{ia}$,  we obtain a formally identical localization formula for the Witten index:
 \be
I[q]  =   ( q - q^{-1})^{1-N} \sum_{z_*} {\rm sgn} \det   (\pa_{z^a}\pa_{z^b} \F (z_*) )q^{ -2 J_3 (z_*)},\label{indIfp}\\
\ee
only now the fixed points are collinear  solutions of the Denef equations (\ref{Denefeqs})
\be 
U_a (z_*)= f_a + \sum_{b\neq a} \frac{\kappa_{ab}}{2|z_{ab}^*| } =0.\label{collDenef} 
\ee
These are extrema of the Coulomb potential
\be
\F = \half \sum_{a<b} \k_{ab} {\rm sgn} z_{ab}  \ln  | z_{ab} | + \sum_a f_a z_{a}.\\
\ee
The consistency conditions in (\ref{consistemcycond}) now coincide and guarantee that the sign factor in (\ref{indIfp}) is well-defined

Let us compare this result to the literature.
Manschot, Pioline and Sen \cite{Manschot:2010qz,Manschot:2011xc,Manschot:2012rx} obtained a fixed point expression for the ground state character $\O_{\rm MPS}[q]$ (see \ref{OmMPS}) which is related to our expression for the Witten index as
\be 
I[q] 
= \O_{\rm MPS} [-q^{-1} ].
\ee
This equality is expected from the comments below (\ref{OmMPS}): the evaluation  at $-q^{-1}$ accounts for the different choice of fermion parity and sign choice in the definition of the index. Some   states  would contribute differently to $\O_{\rm MPS} [-q^{-1}] $ and $I[q]$, namely the pure-Higgs states in representations  $(0, \tilde j)_0$, but these are not visible in the Coulomb branch analysis.

The expression for $\O_{\rm MPS} [q]$ was first  obtained \cite{Manschot:2010qz,Manschot:2011xc} through a conjectured relation with the equivariant volume on moduli space, 
and this relation was subsequently proven in \cite{Kim:2011sc}. Our localization  derivation is similar to that given in \cite{Ohta:2015fpe} though in our view it  resolves some issues with that computation\footnote{In particular, our path integral   twisted  by $J_3 + \tilde J_3$  preserves some supersymmetry,   while a path integral twisted by $J_3$ as in \cite{Ohta:2015fpe} does not and  therefore doesn't obviously  localize.}.

\section{Evaluating the indices }\label{Seceval}
\begin{figure}
\begin{center}
\includegraphics[height=150pt]{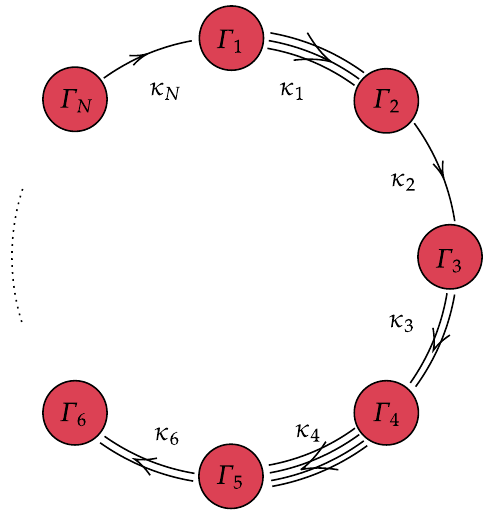}\label{Figcyclicquiver}
\end{center}
\caption{An abelian cyclic quiver:  each node represents a single D-brane center.}
\end{figure}

In this section we turn to the evaluation of our fixed-point formulae (\ref{indfp},\ref{indIfp})  for abelian cyclic quivers with $N$-centers (see Figure \ref{Figcyclicquiver}), in the scaling regime where the DSZ inner products satisfy the inequalities (\ref{Nineqs}). We will discuss both  the Witten index and the superconformal index and find a relation between the two in a certain large-$N$ regime.

To shorten formulas we introduce the  notation 
\be 
Z_a := z_{a, a+1},\qquad  Z_N := z_{N1}; \qquad \k_a :=  \k_{a, a+1},\qquad  \k_N := z_{N1}, \qquad a = 1 , \ldots, N-1
\ee
The parameters $\k_a$ can  be taken to be all positive as we will do in the following. 
\subsection{Witten index in the scaling regime} 
The collinear Denef equations  read in this case
\bea 
{\k_a \over |Z_a|} - {\k_{a+1} \over |Z_{a+1}|} &=& f_{a+1}, \qquad a = 1, \ldots , N\\
Z_1 + Z_2 + \ldots + Z_N &=& 0
\eea
While we refer to \cite{Manschot:2012rx} for the systematics of solving these equations, we want to recall here that besides regular solutions where the centers are separated and the contribution to the index formula is well-defined, due to the inequalities (\ref{Nineqs}) there also exist scaling solutions where the centers come arbitrarily close together. 
In this limit  we can set $f_a \to 0 $ and the collinear Denef equations (\ref{collDenef}) reduce to
\bea 
{\k_1 \over |Z_1|} = {\k_{2} \over |Z_{2}|} = \ldots &=&  {\k_N \over |Z_N|}\label{Denefscaling}\\
Z_1 + Z_2 + \ldots + Z_N &=& 0
\eea
These can be rewritten as 
\bea 
|Z_a| &=& { \k_a \over \k_N} |Z_N|, \qquad a = 1, \ldots, N-1\\ 
\left( \sum_{a=1}^N s_a \k_a \right) Z_N &=& 0
\eea
In the specially tuned cases where the $\k_a$ are chosen so that a linear combination with coefficients $\pm 1$ vanishes, there is a one-parameter of solutions (since $Z_N$ is arbitrary), all of which have vanishing $J_3$. For generic $\k_a$ however, as we shall assume here, the only solution is the  configuration $Z_j =0\  \forall j$, which lies on the singularity where the Coulomb branch description ceases to be reliable,  and for which the index formula (\ref{indIfp}) is not well-defined.  

Excluding these singular solutions and denoting by $I[q]$ the contributions to the Witten index formula (\ref{indIfp}) from regular collinear solutions to (\ref{collDenef}), it was observed by Manschot, Pioline and Sen (MPS) that 
the result cannot be the full answer. First of all, while the index should be a Laurent polynomial (and moreover be an $su(2)_{\rm diag}$ character), the expression  (\ref{indIfp}) in general is not. Furthermore, comparison with more microscopic formulas obtained from the Higgs branch shows that $I[q]$ misses the  `pure-Higgs' states already mentioned above.
The interpretation proposed by MPS is that these properties signify   that $I[q]$ misses contributions from scaling solutions and  that the total
index should include an extra scaling contribution:
\be 
I_{\rm total}[q] = I[q] + I_{\rm scaling}[q].
\ee
In particular, the scaling part $ I_{\rm scaling}[q]$ contains 
a set of `minimal' modification terms necessary to turn the total index into a Laurent polynomial, as well as contributions from the pure-Higgs states.
The missing scaling contribution $I_{\rm scaling}[q]$ can in principle be obtained from the exact microscopic Jeffrey-Kirwan  residue formula for $I_{\rm total}[q]$ derived in the full quiver theory by Hori and Yi \cite{Hori:2014tda}. The result for 
$ I_{\rm scaling}[q]$ was, at least for a subset of  quiver systems (including the abelian cyclic ones of interest), rewritten in a suggestive 
form by Beaujard, Mondal and Pioline in \cite{Beaujard:2021fsk}.   Below we will relate their formula to our expression for the superconformal index in a certain large-$N$ regime.
 
\subsection{Evaluating the superconformal index}
Let us now turn to the evaluation of the superconformal index.
 The conformal localization equations (\ref{confdenefeqs}) for abelian cyclic quivers read
\bea 
{\k_{1}\over |Z_{1}|} \left( 1 + {\o \over Z_{1}}\right) = { \k_{2}\over  | Z_{2}| } \left( 1 + {\o \over Z_{2}}\right) = \ldots = {  \k_{N}\over  | Z_{N} | } \left( 1 + {\o \over Z_{N}}\right) &:=& \g \label{confeqscyc1}\\
Z_1 + Z_2 + \ldots + Z_N &=& 0
\label{confeqscyc2}
\eea
These form a deformation of the scaling Denef equations (\ref{Denefscaling}) which allows for the existence of regular solutions as we shall presently see.
We mentioned before that the strictly positive parameter $\o$  should be spurious and drop out of the index. Indeed, $\o$  can be absorbed in a rescaling of the coordinates (and of the parameter $\g$), which does not alter the index as it  depends only on the ordering of the brane positions. For definiteness we set
\be 
\o =1
\ee
in the rest of this section.

It will be useful to introduce  the sign factors
\be s_a = {\rm sgn}\, Z_a,\qquad  a =1, \ldots, N .\label{signdefs}\ee
To solve the equations (\ref{confeqscyc1},\ref{confeqscyc2}), we follow a  method  similar to \cite{Manschot:2012rx}: we first solve the quadratic equations (\ref{confeqscyc1}) yielding $Z_a$ in terms of $\g$:
\be 
Z_a = {s_a \k_a \over 2 \g} \left( 1 + t_a \sqrt{ 1 + {4  s_a \g \over \k_a}}\right).\label{quadrsol}
\ee
where $t_a = \pm 1$ are further sign choices determining the branches of the quadratic roots. Subsequently, for each choice of $s_a, t_a$, we use (\ref{confeqscyc2}) to solve (in practice numerically) for $\g$. The solution is self-consistent only if the sign of $Z_a$ is indeed $s_a$, or  equivalently if
\be 
 {1\over  \g} \left( 1 + t_a \sqrt{ 1 + {4 s_a \g \over  \k_a}}\right) \geq 0 \qquad \forall a = 1, \ldots, N.
\ee
Selecting the solutions for which this is satisfied, we obtain a collection of solutions $\{ Z_* \}$ which  we should  substitute  in (\ref{indfp}):
\be
\tilde I[q] =  ( q - q^{-1})^{1-N} \sum_{ \{ Z_* \} }  \s( Z_*) q^{-2 J_3 (Z_*) }
\ee
where
\bea
\s( Z_*) &=& {\rm sgn} \det \left({\pa^2 \over \pa_{Z_a Z_b}} \tilde \F\right)_{|Z_*}\label{signsc}\\
J_3 (Z_*)&=& \half  \sum_{a=1}^N \k_{a}  {\rm sgn}\, Z_a^*.
\eea

Let us illustrate this  in the simplest case of 3 centers.  
The sign factor (\ref{signsc}) is in this case given by
\be
s (Z) = {\rm sgn} \left( {\k_1 \k_2 (Z_{1}+2)(Z_{2}+2)\over |Z_{1}|^3 |Z_{2}|^3} +  {\k_1\k_3 (Z_{1}+2)(Z_{3}+2)\over |Z_{1}|^3 |Z_{3}|^3}  +  {\k_2 \k_3 (Z_{2}+2)(Z_{3}+2)\over |Z_{2}|^3 |Z_{3}|^3}  \right).
\ee

The simplest situation occurs when all intersection products are the same,  $\k_1=\k_2=\k_3 \equiv \k$. Then one finds that there are three fixed points:
\be 
(Z_1^*,Z_2^*,Z_3^*) = \left(-{5\over 2},-{5\over 2},5 \right),\left(-{5\over 2},5,-{5\over 2}  \right),\left(5,-{5\over 2},-{5\over 2}  \right).
\ee
One checks that $s(Z_*)=- 1$ in all three cases and
\be 
\tilde I[q] = - {3  q^{\k}   \over (q - q^{-1})^{2} }.
\ee 

In less symmetric situations, the solutions to the fixed point equations can be found numerically as outlined above. We tabulate some sample results in Table \ref{Table3cen}
\begin{table}	\begin{center}
	\begin{tabular}{ |c|c|c|c| c|} \hline
	  $ (\k_1,\k_2,\k_3)$ & $(Z_1^*,Z_2^*,Z_3^*)$ &$ (q-q^{-1})^2  \tilde I[q]$& $(q-q^{-1})^2   \O_{\rm uneq} [-q^{-1}]$\\ \hline  \hline 
	   $(4,5,6)$ & 	\begin{tabular}{c}$(-3.4, -4.7 , 8.1),$ \\$(4.7, -3.4, -1.3)$   \end{tabular} & $-q^3+q^7 $  
		&  $-q^3-q^5-q^7$ \\ \hline 
		    $(4,4,6)$ & 	$(-5.5, -5.5, 11)$ & $-q^2 $  
		   &  $q^2+2 q^6$ \\ \hline 
		     $(82, 102, 113)$ & 	\begin{tabular}{c}$(-2.8, -4.1, 6.9),$ \\$(-1.5 , 6.3 , -4.8 )$  \\ 
		     	$(-1.3, 8.1, -6.8  )$  \\ 	$(4.6, -3.2 , -1.4  )$   \end{tabular} & $-q^{71} +q^{133} $  
		   &  $-q^{71}-q^{93}  -q^{133}$ \\ \hline 
\end{tabular}\end{center}\label{Table3cen}\caption{Some sample results for the superconformal index for the 3-center quiver. In the last column we compare with the quantity $\O_{\rm uneq}$ defined below in (\ref{Omuneq}).} \end{table}
We checked that in all cases the conditions (\ref{consistemcycond}) are indeed satisfied.
For comparison, we have also tabulated the values of the quantity $ \O_{\rm uneq} [-q^{-1}]$ to be discussed in Section \ref{SecPiol} below.

\subsection{Large-$N$ solutions}\label{SeclargeNsols}
 We should keep in mind that, while these results provide the exact index for the model defined by the Lagrangian (\ref{Lagcomp}), we cannot expect to trust them  to  accurately capture the actual D-brane index in general, due to the   neglected corrections. As we discussed in Section \ref{Seccorrs}, we would trust a contribution from a given fixed point only when \be |Z^*_a|\gg 1,\ \forall a, \label{trustfp}\ee i.e. it lies sufficiently far from the singular locus where the Coulomb and Higgs branches meet and corrections become important.  It is clear from the above examples that   (\ref{trustfp}) is generically not met. The situation improves when we allow the number of centers $N$ (which is also the rank of the quiver gauge group) to grow large: we will see that there is a class of fixed points where $|Z^*_a|$ grow linearly with $N$ and for which the superconformal approximation should be reliable.  
 In the next section we will establish that these contributions are in precise agreement with a scaling contribution to  the Witten index. We should note before proceeding that our large-$N$ limit differs from the one taken in \cite{Anninos:2016szt}, where the DSZ inner products $|\k_{ab}|$ were taken to grow large for fixed number of centers.
 
Taking a  limit where the number of centers $N$ becomes large,  we aim to identify  a class of quivers and of solutions to (\ref{confeqscyc1}) where 
\be
|Z^*_a| = \calo (N^1)\  \forall a,  \qquad {\rm  and\ therefore\ }|\g| = \calo (N^{-1} ).\label{largeNass}
\ee 
In fact $\g $ must   be positive as we see from (\ref{quadrsol}).  Inspection  of (\ref{quadrsol}) also shows that such behaviour can only happen for  solutions (\ref{quadrsol}) where all  sign choices $t_a$ are chosen to be +1.
Under these assumptions, to be checked for self-consistency in the end, we can approximate (\ref{quadrsol}) as
 \be 
 Z_a = {s_a \k_a \over \g } + 1 + \ldots
 \ee
 Solving for $\g$ from (\ref{confeqscyc2}) leads to
 \be 
 \g = - {\sum_a s_a \k_a \over N}+ \ldots .
 \ee
 We see that  the assumption (\ref{largeNass}) is self-consistently obeyed   if the $\k_a$ are taken  not to grow with $N$ and 
 \be 
 J_3 (Z_* ) = \half \sum_a s_a^* \k_a  = \calo (1) \qquad {\rm as \ } N\to \infty .\label{J3O1}
 \ee
That is,  we consider  collinear solutions whose angular  momentum   doesn't grow with $N$.   This places some constraints on the $\k_a$ and on the signs $s_a$, which must be chosen sufficiently alternating to allow for cancellations to happen. For example, if all $\k_a$ are equal, we must take the fraction of positive $s_a$ to approach $1/2$ in the large-$N$ limit.   In particular, to satisfy (\ref{J3O1})  the signs $s_a$ cannot all be the same.
 The resulting fixed points are
 \be 
\boxed{  Z^*_a =  - N {s_a \k_a \over  \sum_b   s_b \k_b} + 1 + \calo( N^{-1} )}\label{largeNsolsc}
 \ee 
 and grow indeed as $\calo (N^1)$ in the regime (\ref{J3O1}). Since as we saw above $\g$ has to be positive, we must also have
 \be 
  J_3 (Z^*) <0
 \ee
 which is consistent with the fact that our index was defined in Section \ref{Secind}  to receive contributions from chiral primary states  with lowest angular momentum weights.
 One also shows that for this class of solutions the sign factor simplifies to
 \be 
 \s (Z_*) = \prod_{a=1}^N s_a \ee
  Putting all this together we obtain from these configurations a large-$N$, finite $J_3$ contribution to the index of the form
 \be
\boxed{ \tilde  I[q]_{| N \gg 1, J_3 = \calo (1)} = (q - q^{-1})^{1- N} \sum_{\footnotesize \begin{array}{cc}s_a = \pm 1\\ \sum_a s_ a\k_a < 0\\ \sum_a s_a \k_a = \calo (1) \end{array}} \left(\prod_b s_b\right) q^{ - \sum_a  s_a \k_a}}\label{largeNindsc}
 \ee

\subsection{Relation to microscopic  scaling index}\label{SecPiol}

We will now find a precise relation of this large-$N$ contribution to the superconformal index   with with Beaujard, Mondal and Pioline's  \cite{Beaujard:2021fsk} exact localization formula for the scaling part of the protected spin character $\O_{\rm scaling}[q]$. The latter formula was obtained by rewriting the microscopic Higgs branch expression (leading to the Jeffery-Kirwan residue formula for the   index) \cite{Hori:2014tda} in a way that   suggests a Coulomb branch interpretation. Their expression for the index again localizes on collinear configurations where the positions of the centers satisfy another deformation of the scaling Denef equations (\ref{Denefscaling}), namely
\bea 
{\k_1 \over |Z_1  -  { {\rm Im} \theta \over  \b}  |} = {\k_{2} \over |Z_{2}-  { {\rm Im} \theta \over  \b} |} = \ldots &=&  {\k_N \over |Z_N-  { {\rm Im} \theta \over  \b} |}\nonu
Z_1 + Z_2 + \ldots + Z_N &=& 0\label{BMPscaling}
\eea
These equations   were derived upon allowing the chemical potential $\theta$ (recall that $q := e^{i \theta}$) to acquire an imaginary part. It plays the role of a `real mass' for the chiral multiplets which makes the path-integral expression for the index well-defined \cite{Hori:2014tda}. 
A first observation is that    for $|Z_a | \gg  {{\rm Im} \theta \over  \b}$ the equations (\ref{BMPscaling}) agree to first subleading order
with the conformal localization equations (\ref{confeqscyc1}) upon identifying
\be 
\o= { {\rm Im} \theta \over  \b}.
\ee
Let us unpack and generalize this observation a bit. For general scaling quivers the localization equations  of \cite{Beaujard:2021fsk} can be derived from a Coulomb potential
\be 
\F_{\rm scaling} = \half \sum_{a<b} \k_{ab} {\rm sgn} \left( z_{ab} - { {\rm Im} \theta \over \b} \tilde J_3 (\f_{ab} ) \right) 
\ln  \left|  z_{ab} - { {\rm Im} \theta \over \b} \tilde J_3 (\f_{ab} )\right|,
\ee
where $\tilde J_3 (\f_{ab} )$ is the R-charge of the  chiral multiplets from strings stretched between the $a$ and $b$ centers  (defined such that $\tilde J_3 (\f_{ba} )= -\tilde J_3 (\f_{ab} )$). Expanding this for $|z_{ab}| \gg |  { {\rm Im} \theta} \tilde J_3 (\f_{ab} )|/\b $ to first subleading order
and comparing with  the Coulomb potential $\tilde \F$  for the superconformal localization equations (\ref{tildeF}) one sees that they agree provided that we identify
\be 
K =  \sum_{a<b} {\k_{ab} \tilde J_3 (\f_{ab} ) \over 2 |z_{ab}|},
\ee
or equivalently if the $\tilde J_3 (\f_{ab} )$ can be chosen (up to an irrelevant rescaling) to be
\be 
 \tilde J_3 (\f_{ab} ) \sim {\rm sgn } \k_{ab},\ee
 which is indeed the case for abelian cyclic quivers.

We will now see that a class of large-$N$  solutions to (\ref{BMPscaling}) and their contributions to the Witten index match with (\ref{largeNsolsc}, \ref{largeNindsc}). The parameter ${\rm Im} \theta$ is again a spurious parameter as the index  will depend only on the ordering of the D-brane positions on the $z$-axis, and only the sign of ${\rm Im} \theta$ (which we assume positive here)  actually matters.  We can therefore again set 
\be 
\o = { {\rm Im} \theta \over  \b}=1
\ee
by rescaling the $Z_a$.   
Introducing the sign factors
\be 
\tilde s_a = {\rm sgn} \left( Z_a  -1  \right)
\ee
the general solution to (\ref{BMPscaling}) is
\be
\left| {Z_a } - 1 \right| = -  N {  \k_ a \over \sum_b \tilde s_b \k_b}.\label{gensolPioline}
\ee
The sign choices $\tilde s_a$ are constrained only by the requirement that the right-hand side of this equation is positive, i.e. that
\be 
{\rm sgn}\left( \sum_b \tilde s_b \k_b \right)<0 ,
\ee
which leaves $2^{N-1}$ allowed sign choices out of $2^N$ possibilities.

It was shown in \cite{Beaujard:2021fsk} that the contribution of these fixed points to the protected spin character is qualitatively different in the case where all the signs $\tilde s_a$ are equal and in  the case where they are not. Therefore it is convenient to separate  these contributions:
\be 
\O_{\rm scaling} [q] = \O_{\rm same} [q] + \O_{\rm uneq} [q] .
\ee
For the evaluation of $\O_{\rm same}[q]$, which is more involved, we refer to \cite{Beaujard:2021fsk}.
 We focus here only $\O_{\rm uneq}[q] $ since this is the contribution which will match with the superconformal index at large $N$ (recall that our signs $s_a$ in the large-$N$ solution (\ref{largeNsolsc}) could not all be equal). The expression found in \cite{Beaujard:2021fsk}   is
 \be 
\O_{\rm uneq}[q]  = (q - q^{-1})^{1-N} \sum_{\footnotesize \begin{array}{cc}\tilde s_a = \pm 1\\ \sum \tilde s_a \k_a<0\end{array}}\;\hspace{-0.6cm}\vspace{0.3cm}' \ \ \left(\prod_c \tilde s_c \right)(- q)^{  \sum_a \tilde s_a \k_a}  ,\label{Omuneq}
 \ee
  where the prime on  the sum means that we exclude the configuration with all signs $\tilde s_a$ equal.
  
  For example, in the 3-center case the unequal contributing signs are, due to the triangle inequalities, $(\tilde s_1, \tilde s_2, \tilde s_3) = (+,-,-),(-,+,-),(--+) $,  leading to 
  \be 
  \O_{\rm uneq}[q]  = (q - q^{-1})^{-2}\left( (-q)^{- \k_1 + \k_2 + \k_3} +(-q)^{ \k_1 - \k_2 + \k_3} +(-q)^{ \k_1 + \k_2 - \k_3} \right)
  \ee
    This can be compared (as we did in Table \ref{Table3cen}) with the superconformal index; one notes that the latter often misses contributions while also the sign factor $\s (Z_* )$ is typically different.
  
 Similarly to the previous section we can let $N$ grow large and consider sequences of quivers and fixed points where $\sum_b \tilde s_b \k_b$ stays of order one, so that all $Z_a$ are of order $N$. Note that the sign factors can then be replaced by 
  \be
 \tilde s_a = {\rm sgn} (Z_a - 1) = {\rm sgn} Z_ a = s_a.
 \ee
The fixed point solutions (\ref{gensolPioline}) and (\ref{largeNsolsc}) clearly agree in this limit,    and their contributions to the Witten and conformal indices are  related as
\be 
\boxed{ \tilde I[q]_{N \gg 1, J_3 = \calo (1)} =  \O_{\rm uneq}[-q^{-1}]_{N \gg 1, J_3 = \calo (1)}.}\label{scPioline}
\ee
This observation constitutes the main result of this Section. 

\section{Discussion and outlook}\label{Secdisc}
In this work we have derived a fixed-point formula for the superconformal index in quiver mechanics. In a certain large-$N$ regime, where we expect the conformal description to be reliable, we showed in  (\ref{scPioline})  that our expression captures a part of the  microscopic scaling index 
 which was hitherto not visible on the Coulomb branch. This result is suggestive that  conformal symmetry plays a role  in  the scaling BPS   spectrum in the Coulomb branch and it is hoped provide a step  towards  a stringy realization of AdS$_2$/CFT$_1$ duality. Our results raise several questions which require a deeper understanding.

 Firstly, the relation  (\ref{scPioline}) between superconformal and  Witten indices is  somewhat surprising\footnote{ One notable example in the literature where the superconformal and Witten-type indices are related is \cite{Cordova:2015nma}.}.  
 Though both indices are regularizations of the object $ \tr(-1)^{2 \tilde J_3} q^{-2(J_3 + \tilde J_3)}$, they use different Hamiltonians in the Bolzmann factor $e^{-\b H}$  which are not obviously related by a `small deformation' of parameters. 
At the very heuristic level, we recall  that the microscopic computation of $\O_{\rm micro} [q]$ relies on turning on an imaginary part of the chemical potential $\theta$ which plays the role of a  `real mass'  in order to keep the index well-defined. We observe that the effective Hamiltonian $\tilde H (\theta)$ in (\ref{tildeHtheta}) in the superconformal index contains the term $2 (\o + i \theta / \b) J_3$, suggesting that the superconformal index  in a sense turns on a `real mass' regularization in the Coulomb branch. It would also be interesting to understand better how our findings are related to the work \cite{Anninos:2016szt} (see also \cite{Biggs:2023mfn}) which found an emergent  conformal symmetry on the Higgs branch upon averaging over superpotential coefficients, especially since it emerges in a different large-$N$ limit and leads to different scaling dimensions of the fields.

Another question which requires clarification is the physical interpretation of the states contributing to $\O_{\rm uneq}[q] $, which we see in the superconformal regime, especially whether they include `pure-Higgs' states. Naively, one would not expect so since, as we discussed in Section \ref{Secreps}, pure-Higgs-like states would give rise to pole or constant terms in the Laurent series of $\tilde I [q]$, while our (\ref{largeNindsc}) does not contain such terms. Therefore, it seems plausible that the superconformal contribution captures the large-$N$ limit of the `minimal modification terms' of  \cite{Manschot:2012rx}.

 We conclude with some interesting directions for future research:
\begin{itemize}
\item While the current work has mostly focused on the simple class of abelian cyclic quivers, it would be interesting to understand the role of conformal symmetry in more general scaling quivers. 
	\item While we have seen that part the of scaling index, namely $\O_{\rm uneq} [q]$, is captured by a superconformal index, it is unclear if the remaining part $\O_{\rm same} [q]$ might also have a conformal interpretation. For this it would be interesting to investigate the possible emergence of conformal symmetry in the full quiver theory in a large-$N$ limit of the type studied here.
	\item Since the deep scaling limit on the Coulomb branch corresponds to an AdS$_2$ limit of the corresponding 4D multi-centered supergravity  solutions \cite{Bates:2003vx}, it would be interesting to see if the superconformal index can be computed directly in supergravity, and if the contributing configurations obey deformed Denef equations similar to (\ref{confdenefeqs}). Recent progress in understanding multi-centered contributions to indices was made in \cite{Boruch:2025biv}.
\end{itemize}

\section*{Acknowledgements}
It is a pleasure to thank Dionysis Anninos, Jan Boruch,  Elias Kiritsis, Shiraz Minwalla and Boris Pioline  for useful discussions and correspondence, and Dieter Van den Bleeken and Tom\'a\v{s} Proch\'azka for initial collaboration on this project.  The research of A.D.M., J.R. and P.R. was  co-funded by the European Union and supported by the Czech Ministry of 
Education, Youth and Sports  (Project FORTE CZ.02.01.01/00/22\_008/0004632). M.M. and G.S. were supported by the Scientific and Technological Research Council of Turkey (TÜBİTAK) under the grant with number 123N952. This collaboration benefited from travel support under the mobility grant TÜBİTAK 24-14. 

\begin{appendix}
	\section{Spinor index conventions}\label{Appferm}
In this Appendix we spell out our spinor index conventions. Spinor indices $\a, \tilde \b, \ \a,\ \tilde \b  \in \{+, -\}$ are raised and lowered using antisymmetric $\e$-tensors. Index contractions follow a `southwest-northeast' convention:
	\bea \l^{\a \tilde \a}  &=& \l_\b^{\ \ \tilde \a} \e^{\b\a}, \qquad \l_\a^{\ \ \tilde \a} = \e_{\a\b}\l^{\b \tilde \a},\qquad 
	\l^{\a \tilde \a}  = \l^\a_{\ \ \tilde \b} \tilde \e^{\tilde \b \tilde \a}, \qquad \l^\a_{\ \ \tilde \a} = \e_{\tilde \a\tilde \b}\l^{\a \tilde \b}, \\
	\l \n &:=& \l_{\a \tilde \a} \n^{ \a \tilde \a} ,
	\eea
	where $\e^{+-} = \e_{+-}=\tilde \e^{+-} =\tilde  \e_{+-}=1$. The reality conditions on the spinors are
	 \be 
	(\l^{\a \tilde \b})^* = \a \tilde \b \l^{ -\a, -\tilde \b}
	\ee
	 and conjugation is understood to invert the ordering in  products of spinors. 
	Pauli matrices $\s_i, \tilde \s_i$ act on untilded and tilded indices respectively and have components
	\be 
	(\s_i)_\a^{\ \b} = 	(\tilde \s_i)_{\tilde \a}^{\ \tilde \b}  = \left\{\left( \begin{array}{cc} 1&0\\0&-1 \end{array} \right),\left( \begin{array}{cc} 0&-i\\i&0 \end{array} \right) ,\left( \begin{array}{cc} 0&1\\1&0 \end{array} \right)  \right\}.
		\ee
		Note that the Pauli matrices satisfy
	$$ (\s_i)^\a_{\ \b } = (\s_i)_\b^{\ \a} = { (\s_i)_\a^{\ \b}}^*.$$
	Here are some examples of index-contracted expressions involving Pauli matrices:
	\be 
	\l \s_i \n := \l_{\a \tilde \a} (\s_i)^\a_{\ \b} \l^{\b \tilde \a} , \qquad 	\l \tilde \s_i \n := \l_{\a \tilde \a} (\tilde \s_i)^{\tilde \a}_{\ \tilde \b} \l^{\a \tilde \b}.
	\ee

	\section{Dirac Bracket analysis}\label{AppDB}
In this Appendix we derive the Dirac brackets between the canonical variables.	The conjugate momenta are 
	\begin{align}
		p_{ia}&=\frac{\partial L}{\partial\dot{x}^{ia}}=G_{ab}\dot{x}^{ib}-A_{ia}-\frac{1}{4}\epsilon_{ijk}\partial_{ja}G_{bc}\lambda^b\sigma_k\lambda^c;\\
		\pi_{a\alpha\tilde{\alpha}}&=\frac{\partial L}{\partial\dot{\lambda}^{a\alpha\tilde{\alpha}}}=\frac{i}{2}G_{ab}\epsilon_{\alpha\beta}\epsilon_{\tilde{\alpha}\tilde{\beta}}\lambda^{b\beta\tilde{\beta}}.\label{fermmom}
	\end{align}
	We impose the standard graded-Poisson brackets  
	\begin{align}
		\{x^{ia},p_{jb}\}_P
		&=\delta^i_j\delta^a_b,\\
		\{\lambda^{a\alpha\tilde{\alpha}},\pi_{b\beta\tilde{\beta}}\}_P
		&=\delta^a_b\delta^\alpha_\beta\delta^{\tilde{\alpha}}_{\tilde{\beta}}.
	\end{align}
	We notice that (\ref{fermmom}) leads to the constraints
	\be 
	\psi_{a\alpha\tilde{\alpha}}:=\pi_{a\alpha\tilde{\alpha}}-\frac{i}{2}G_{ab}\epsilon_{\alpha\beta}\epsilon_{\tilde{\alpha}\tilde{\beta}}\lambda^{b\beta\tilde{\beta}}
	\ee
These  constraints are  second-class as follows from the Poisson brackets 
	\begin{align}
		C_{a\alpha\tilde{\alpha}b\beta\tilde{\beta}}&: =\{\psi_{a\alpha\tilde{\alpha}},\psi_{b\beta\tilde{\beta}}\}_P
		 =-iG_{ab}\epsilon_{\alpha\beta}\epsilon_{\tilde{\alpha}\tilde{\beta}};\\
		C^{a\alpha\tilde{\alpha}b\beta\tilde{\beta}}&=iG^{ab}\epsilon^{\alpha\beta}\epsilon^{\tilde{\alpha}\tilde{\beta}}
	\end{align}
	We  take them into account using the standard Dirac bracket.	
	The non-vanishing (Poisson) brackets among the constraints and the canonical variables are 
	\begin{align}
		\{\lambda^{a\alpha\tilde{\alpha}},\psi_{b\beta\tilde{\beta}}\}_P
		&=\delta^a_b\delta^\alpha_\beta\delta^{\tilde{\alpha}}_{\tilde{\beta}},\\
		\{\pi^{a\alpha\tilde{\alpha}},\psi_{b\beta\tilde{\beta}}\}_P
		&=-\frac{i}{2}G_{ab}\epsilon_{\alpha\beta}\epsilon_{\tilde{\alpha}\tilde{\beta}},\\
		\{p_{ia},\psi_{b\beta\tilde{\beta}}\}_P
		&=\frac{i}{2}\partial_{ia}G_{bc}\epsilon_{\beta\gamma}
		\epsilon_{\tilde{\beta}\tilde{\gamma}}
		\lambda^{c\gamma\tilde{\gamma}}.
	\end{align}
The Dirac bracket is defined as
	\begin{align}
		\{\quad,\quad\}_D&=\{\quad,\quad\}_P-\{\quad,\psi_{a\alpha\tilde{\alpha}}\}_PC^{a\alpha\tilde{\alpha}b\beta\tilde{\beta}}\{\psi_{b\beta\tilde{\beta}},\quad\}_P.
	\end{align}
	The only brackets that are modified are
	\begin{align}
		\{\lambda^{a\alpha\tilde{\alpha}},\pi_{b\beta\tilde{\beta}}\}_D
		&=\frac{1}{2}\delta^a_b\delta^\alpha_\beta\delta^{\tilde{\alpha}}_{\tilde{\beta}},\\
		\{\lambda^{a\alpha\tilde{\alpha}},p_{ib}\}_D
		&=-\frac{1}{2}G^{ac}\partial_{ib}G_{cd}\lambda^{d\alpha\tilde{\alpha}},
	\end{align}
	which reduces to the expressions given in (\ref{DB})
\be 
		\{\lambda^{a\alpha\tilde{\alpha}},\lambda_{b\beta\tilde{\beta}}\}_D
		= -i\,\delta^a_b \delta^\alpha_\beta \delta^{\tilde{\alpha}}_{\tilde{\beta}}\qquad 
		\{p_{ia},\lambda_b^{\alpha\tilde{\alpha}}\}_D
		= \frac{1}{2}\partial_{ic}G_{ab}\lambda^{c\alpha\tilde{\alpha}}.
\ee
\end{appendix}                                                                                                                                                                       
\bibliographystyle{ytphys}
\bibliography{referencesD21a}
\end{document}